\documentclass{amsart}
\usepackage{amsmath}
\usepackage{amsthm}
\usepackage{amssymb}
\usepackage{graphicx}
\usepackage{epsbox}

\newtheorem{theorem}{Theorem}
\newtheorem{lemma}[theorem]{Lemma}

\newtheorem{proposition}[theorem]{Proposition}
\numberwithin{theorem}{section}

\theoremstyle{definition}

\theoremstyle{remark}
\newtheorem{remark}[theorem]{Remark}

\numberwithin{equation}{section}

\newfont{\germ}{eufm10}

\newcommand{\gehn}{\ensuremath{\mathfrak{g}_n}}

\def\Z{\mathbb Z}
\def\C{\mathbb C}
\def\ot{\otimes}

\def\openone{\leavevmode\hbox{\small1\kern-3.8pt\normalsize1}}

\begin{document}

\title{A quantization of box-ball systems}

\begin{abstract}
An $L$ operator is presented related to an 
infinite dimensional limit of the fusion $R$ matrices for 
$U_q(A^{(1)}_{n-1})$ and $U_q(D^{(1)}_n)$.
It is factorized into the local propagation operators 
which quantize the deterministic dynamics of particles and antiparticles 
in the soliton cellular automata 
known as the box-ball systems and their generalizations.
Some properties of the dynamical amplitudes are also investigated.
\end{abstract}

\author{R. Inoue}
\address{Research Institute for Mathematical Sciences, 
Kyoto University, Kyoto 606-8502, Japan}
\email{reiiy@kurims.kyoto-u.ac.jp}

\author{A. Kuniba}
\address{Institute of Physics, University of Tokyo, Tokyo 153-8902, Japan}
\email{atsuo@gokutan.c.u-tokyo.ac.jp}

\author{M. Okado}
\address{Division of Mathematical Science,
    Graduate School of Engineering Science,
Osaka University, Osaka 560-8531, Japan}
\email{okado@sigmath.es.osaka-u.ac.jp}

\maketitle

\section{Introduction}\label{sec:intro}

The discovery of the box-ball systems \cite{TS,T,TTMS} and 
their connection to the crystal basis theory \cite{HKT1,HHIKTT,FOY} 
has led to a new parallelism across the integrable systems 
of three origins, quantum, ultradiscrete and classical \cite{KOTY2}.
They are a class of two dimensional vertex models in statistical mechanics,
one dimensional soliton cellular automata and 
discrete soliton equations.
The fundamental objects that govern the local dynamics 
in these systems are the triad of 
quantum $R$, combinatorial $R$ and tropical $R$,
all satisfying the Yang-Baxter equation.
They are a finite dimensional matrix, a bijection 
among finite sets and a birational map, which are 
characterized as the intertwiners
of $U_q$ modules, crystals and geometric crystals, respectively.
The box-ball systems $(\gehn=A^{(1)}_{n-1})$
and their generalizations to the $\gehn$ automata \cite{HKT1,HKOTY}
are associated with the combinatorial $R$, which 
arises both as the $q \rightarrow 0$ limit of the quantum $R$ and 
as the ultradiscretization of the tropical $R$ \cite{KOTY1}.

An interesting feature in these automata is the factorization of 
time evolution into a product of propagation operators of 
particles and antiparticles with fixed color \cite{HKT3,KTT}.
This is a consequence of the factorization of 
the combinatorial $R$ shown in \cite{HKT2}.
Our aim in this paper is to elucidate a similar factorization 
for the relevant quantum $R$, and thereby to launch an
integrable quantization of the deterministic dynamics 
of particles and antiparticles 
in the generalized box-ball systems.

To illustrate the idea, consider 
for example the quantum affine algebra $U_q(A^{(1)}_{n-1})$ 
and its irreducible finite dimensional 
representation $V_m$ of $m$ fold symmetric tensors.
The quantum $R$ matrix for $V_m \ot V_1$ (\ref{eqa:Rm1w}) 
gives rise to the commuting transfer matrix $T_m(z)$ acting on 
$\cdots \ot V_1 \ot V_1 \ot \cdots$, which reduces, 
at $q=0$, to the time evolution of the box-ball system 
with capacity $m$ carrier \cite{TM}.
One can naturally extract an $L$ operator, a Weyl algebra valued matrix, 
{}from the $m \rightarrow \infty$ limit of 
the $R$ matrix in the vicinity of the lowest weight vector.
See (\ref{eqa:ex23}) and (\ref{eqa:ex4}) for example.
More general $L$ operators can be constructed 
similarly corresponding to the $m$ generic situation.
The limit considered here is motivated by the box-ball systems and 
has a special feature in that the resulting $L$
admits the factorization as in Proposition \ref{pra:factor}.
Each operator $K_i$ appearing there encodes the 
amplitudes for a local propagation of  
color $i$ particles as depicted in Fig. \ref{fig:Ki}.
At $q=0$, it reduces to the deterministic dynamics
in the box-ball system \cite{T}.

Sections \ref{subsec:R}--\ref{subsec:qbbs3} are devoted to 
an exposition of these observations.
Sections \ref{subsec:norm} and \ref{subsec:ba} 
are concerned with some properties of the dynamical amplitudes 
and the implication of the Bethe ansatz, respectively.
In section \ref{sec:D} we establish parallel results 
on $D^{(1)}_n$ case.
The calculation of the fusion 
$R \in \text{End}(V_m \ot V_1)$ is more involved than $A^{(1)}_{n-1}$. 
It is done in the limit 
$m \rightarrow \infty$ in appendix \ref{appD:W}.
The $L$ operator is given in section \ref{subsec:Ld}
and factorized in section \ref{subsec:facLd}.
The propagation operators describe the amplitudes of 
pair creation and annihilation of particles and 
antiparticles as depicted in Fig. \ref{fig:Kmu}.
A quantized $D^{(1)}_n$ automaton is presented 
in section \ref{subsec:dbbs} with 
a few basic properties.

The fusion construction of the $R$ matrices and 
their matrix elements for $A^{(1)}_{n-1}$ 
given in section \ref{sec:A} are not new. 
They have been included for the sake of self-containment.
The content of this paper may be regarded as 
a generalization of the one in \cite{HKT2} for $q=0$.
It will be interesting to investigate 
the present results in the light of the works \cite{KT,KR,S}.

\section{\mathversion{bold}$A^{(1)}_{n-1}$ case}\label{sec:A}
\subsection{\mathversion{bold}
$R$ matrix $R(z)$ and its fusion $R^{(m,1)}(z)$}
\label{subsec:R}
We recall the standard fusion construction \cite{KRS}.
Let $V=\C v_1 \oplus \cdots \oplus \C v_n$ 
be the vector representation of 
the quantum affine algebra  $U_q=U_q(A^{(1)}_{n-1})$ 
without the derivation operator.
Here $v_1$ is the highest weight vector and 
our convention of the coproduct is 
$\Delta(e_i) = e_i\ot 1 + t_i\ot e_i,
\Delta(f_i) = f_i\ot t^{-1}_i + 1 \ot f_i$ 
for the Chevalley generators.
The $R$ matrix $R(z) \in {\rm End}(V\ot V)$ reads
\begin{equation}\label{eqa:r}
\begin{split}
&R(z) = a(z)\sum_iE_{ii}\ot E_{ii} +
b(z)\sum_{i\neq j}E_{ii}\ot E_{jj} 
+ c(z)\left(z\sum_{i<j}+\sum_{i>j}\right)
E_{ji}\ot E_{ij},\\
&a(z) = 1-q^2z,\quad b(z) = q(1-z),\quad c(z) = 1-q^2,
\end{split}
\end{equation}
where $E_{ij}$ is the matrix unit acting as 
$E_{ij}v_k = \delta_{jk}v_i$.
It satisfies the Yang-Baxter equation
$R_{23}(z'/z)R_{13}(z')R_{12}(z)
= R_{12}(z)R_{13}(z')R_{23}(z'/z)$.
The matrix ${\check R}(z)=PR(z)$ commutes with $\Delta(U_q)$,
where $P$ denotes the transposition of the components.

Let $V_m$ be the irreducible $U_q$ module 
spanned by the $m$ fold $q-$symmetric tensors.
We take $V_1 = V$ and realize the space $V_m$ as the quotient
$V^{\otimes m}/A$,
where $A= \sum_j V^{\otimes j} \ot {\rm Im}PR(q^{-2}) \ot V^{\ot m-2-j}$.
It is easy to see ${\rm Im}PR(q^{-2}) = {\rm Ker}PR(q^{2}) 
= \bigoplus_{i<j}\C(v_i\ot v_j - q v_j \ot v_i)$.
For $n \ge i_1 \ge \cdots \ge i_m \ge 1$, 
we write the vector 
$(v_{i_1}\ot \cdots \ot v_{i_m} \mod A) \in V_m$ as 
$x=[x_1,\ldots, x_n]$, where $x_i$ is the number of the letter 
$i$ in the sequence $i_1, \ldots, i_m$.
Thus, $x_i \in \Z_{\ge 0}$ and $x_1 + \cdots + x_n = m$ holds.

Due to the Yang-Baxter equation, the operator 
\begin{equation}\label{eqd:Rcomp}
\frac{R_{1,m+1}(zq^{m-1})R_{2,m+1}(zq^{m-3})\cdots R_{m,m+1}(zq^{-m+1})}
{a(zq^{m-3})a(zq^{m-5}) \cdots a(zq^{-m+1})}
\end{equation}
can be restricted to ${\rm End}(V_m \ot V)$. 
As a result we get an $m$ by 1 fusion $R$ matrix
$R^{(m,1)}(z) \in {\rm End}(V_m \ot V)$, which reads explicitly as
\begin{align}
&R^{(m,1)}(z)(x\ot v_j) = \sum_k w_{j k}[x \vert y] (y \ot v_k),
\label{eqa:Rm1w}\\
&w_{j k}[x \vert y] = 
\begin{cases}
q^{m-x_k}-q^{x_k+1}z & j=k\\
(1-q^{2x_k})q^{x_{k+1}+x_{k+2}+\cdots+ x_{j-1}}z & j>k\\
(1-q^{2x_k})q^{m-(x_j+x_{j+1}+\cdots+x_{k})} & j<k.
\end{cases}
\label{eqa:element}
\end{align}
It is customary to attach the matrix element $w_{j k}[x \vert y]$ with 
a diagram like Fig. \ref{fig:wjk}.
\begin{figure}
\caption{Diagram for $w_{j k}[x \vert y]$}\label{fig:wjk}
\setlength{\unitlength}{1.3mm}
\begin{picture}(50,17)(-5,18)

\put(10,24.8){\vector(1,0){10}}
\put(10,25.2){\vector(1,0){10}}
\put(15,30){\vector(0,-1){10}}
\put(7.8,24.3){$x$}\put(21,24.3){$y$}
\put(14.2,31.6){$j$}\put(14.2,16.7){$k$}
\put(-10,24){$w_{j k}[x \vert y] = $}

\end{picture}
\end{figure}

\noindent
Here $y=[y_i]$ is specified by the weight conservation as 
\begin{equation}\label{eqa:y}
y_i = x_i + \delta_{i j}- \delta_{i k}
\end{equation}
in terms of $x, j$ and $k$.
At $q=0$, the matrix element $w_{j k}[x \vert y]$ 
is nonzero if and only if 
$x \ot v_j \simeq v_k \ot y$ 
in the combinatorial $R$:  $B_m \ot B_1 \simeq B_1 \ot B_m$, where 
it takes the value $z^H$, with $1-H=$ winding number \cite{NY}.
The fusion $R$ matrix $R^{(m,1)}(z)$ reduces to 
$R(z)$ in (\ref{eqa:r}) for $m=1$, 
and it satisfies the Yang-Baxter equation in ${\rm End}(V_m \ot V \ot V)$:
\begin{equation}\label{eqa:ybe2}
R_{23}(z'/z)R^{(m,1)}_{13}(z')R^{(m,1)}_{12}(z)
= R^{(m,1)}_{12}(z)R^{(m,1)}_{13}(z')
R_{23}(z'/z).
\end{equation}

The $R$ matrix $R^{(1,m)}(z) \in {\rm End }(V \ot V_m)$ is 
similarly obtained as
$R^{(1,m)}(z)(v_j \ot x) 
= \sum_k \bar{w}_{j k}[x \vert y] (v_k \ot y)$, where 
\begin{equation*}
\bar{w}_{j k}[x \vert y] = 
\begin{cases}
q^{m-x_k}-q^{x_k+1}z & j=k\\
(1-q^{2x_k})q^{m-(x_k+x_{k+1}+\cdots+x_{j})} & j>k\\
(1-q^{2x_k})q^{x_{j+1}+x_{j+2}+\cdots+ x_{k-1}}z & j<k.
\end{cases}
\end{equation*}
The inversion relation 
\begin{equation}\label{eqa:inv}
PR^{(1,m)}(z^{-1})PR^{(m,1)}(z) = (1-q^{m+1}z)(1-q^{m+1}z^{-1})\text{Id}
\end{equation}
is valid.

\subsection{\mathversion{bold} $L$ operator $L(z)$}
\label{subsec:L}
Now we extract an $L$ operator $L(z)$ 
{}from a certain limit of $R^{(m,1)}(z)$.
We illustrate the idea along the $n=3$ case.
The 3 by 3 matrix $(w_{ji}[x \vert y])_{1 \le i,j \le 3}$ with 
$y$ chosen as (\ref{eqa:y}) looks as
\begin{equation*}
\begin{pmatrix}
q^{x_2+x_3}-q^{x_1+1}z & (1-q^{2x_1})z & (1-q^{2x_1})q^{x_2}z \\
(1-q^{2x_2})q^{x_3} & q^{x_1+x_3}-q^{x_2+1}z & (1-q^{2x_2})z \\
1-q^{2x_3} & (1-q^{2x_3})q^{x_1} & q^{x_1+x_2}-q^{x_3+1}z
\end{pmatrix}.
\end{equation*} 
Throughout the paper we assume that $\vert q \vert < 1$.
Consider the limit $m \rightarrow \infty$ with 
$x_1$ and $x_2$ kept fixed. 
Namely we take $x_3 \rightarrow \infty$ and stay in the vicinity of 
the lowest weight vector of $V_m$ as $m$ goes to infinity.
The above matrix simplifies to 
\begin{equation}\label{eqa:mat3}
\begin{pmatrix}
-q^{x_1+1}z & (1-q^{2x_1})z & (1-q^{2x_1})q^{x_2}z \\
0 & -q^{x_2+1}z & (1-q^{2x_2})z \\
1 & q^{x_1} & q^{x_1+x_2}
\end{pmatrix}.
\end{equation} 
In the limit, the constraint $x_1+x_2 \le m$ becomes void and 
the vector $x =[x_1,x_2,x_3] \in V_m$ gets effectively labeled as $[x_1,x_2]$
with arbitrary $x_1, x_2 \in \Z_{\ge 0}$.
For generic (nonzero) $x_1$ and $x_2$, 
the $(1,2)$ element $(1-q^{2x_1})z$ in (\ref{eqa:mat3}), for example, 
is the matrix element of the transition 
$[x_1,x_2] \rightarrow [x_1-1,x_2+1]$ in view of (\ref{eqa:y}).
Similarly the $(2,3)$ element 
$(1-q^{2x_2})z$ is the one for $[x_1,x_2] \rightarrow [x_1,x_2-1]$.
Introducing the operator $P_2$ and $Q_2$ that act on 
$[x_1, x_2]$ as $P_2[x_1,x_2]= q^{x_2}[x_1,x_2]$ and  
$Q_2[x_1,x_2]= [x_1,x_2+1]$, the $(2,3)$ element of (\ref{eqa:mat3}) is 
represented as $zQ_2^{-1}(1-P^2_2)$.
With the similar operators $P_1$ and $Q_1$ concerning the 
coordinate $x_1$, the matrix (\ref{eqa:mat3}) is presented as 
\begin{equation}\label{eqa:mat3pq}
\begin{pmatrix}
-zqP_1 & zQ^{-1}_1(1-P_1^2)Q_2 & zQ^{-1}_1(1-P_1^2)P_2 \\
0 & -zqP_2 & zQ^{-1}_2(1-P^2_2) \\
Q_1 & P_1Q_2 & P_1P_2
\end{pmatrix}.
\end{equation}
where operators are all commutative except 
$P_iQ_i = qQ_iP_i$.

Motivated by these observations, we prepare for general $n$ 
the Weyl algebra generated by the pairs
$P_i^{\pm 1}, Q_i^{\pm 1}\; (1 \le i \le n-1)$ 
under the relations 
\begin{equation}\label{eqa:pqcom}
\begin{split}
&Q_iQ_j=Q_jQ_i, \quad P_iP_j=P_jP_i,\quad P_iQ_j = q^{\delta_{ij}}Q_jP_i,\\
&Q_iQ^{-1}_i = Q^{-1}_iQ_i=1, \quad P_iP^{-1}_i =P^{-1}_iP_i=1.
\end{split}
\end{equation}
We actually consider a slight generalization of (\ref{eqa:mat3pq}) 
containing parameters $a_1, \ldots, a_{n-1}$.
Let ${\mathcal A}$ be the subalgebra of the Weyl algebra 
generated by 
\begin{equation}\label{eqa:pqr}
P_i, \;\;  Q_i, \;\;  R_i = Q^{-1}_i(1-a_iP^2_i)\quad 1 \le i \le n-1.
\end{equation}
We also use the subsidiary symbol $P'_i  = -a_iqP_i$.
The previous discussion corresponds to $\forall a_i=1$ case. 
The combination 
$R_i \in {\mathcal A}$ 
introduced here should not be confused with the $R$ matrix.
Then we define the operator $L(z) \in {\mathcal A}\ot {\rm End}(V)$ by
\begin{equation*}
L(z) = \begin{pmatrix}
L_{11}(z) & \cdots & L_{1n}(z)\\
\vdots & \ddots & \vdots\\
L_{n1}(z) & \cdots & L_{nn}(z)
\end{pmatrix},
\end{equation*}
where $L_{ij}(z)\in {\mathcal A}$ is given by
($P_{i,j} = P_iP_{i+1}\cdots P_j$ for $i\le j$)
\begin{equation}\label{eqa:L-elements}
L(z)_{i i} = \begin{cases}
zP'_i & i < n,\\
P_{1, n-1} & i = n,
\end{cases}
\qquad
L(z)_{i j} = \begin{cases}
zR_iP_{i+1,j-1}Q_j & i < j < n,\\
zR_iP_{i+1,n-1} & i < j=n,\\
P_{1,j-1}Q_j &  j<i=n,\\
0 & j<i<n.
\end{cases}
\end{equation}
This is an operator interpretation of 
$w_{ji}[x \vert y]$ (\ref{eqa:element}) in the limit 
$x_n \rightarrow \infty$ deformed with $a_1, \ldots, a_{n-1}$.
See (\ref{eqa:LW}).
For example for $A^{(1)}_{1}$ and $A^{(1)}_{2}$, they read
\begin{equation}\label{eqa:ex23}
L(z) = \begin{pmatrix}
zP'_1 & zR_1\\
Q_1 & P_1
\end{pmatrix},\qquad
L(z) = \begin{pmatrix}
zP'_1 & zR_1Q_2 & zR_1P_2\\
0    & zP'_2   & zR_2\\
Q_1 & P_1Q_2& P_{1,2}
\end{pmatrix}.
\end{equation}
The latter agrees with (\ref{eqa:mat3pq}) when $\forall a_i = 1$. 
For $A^{(1)}_{3}$ one has
\begin{equation}\label{eqa:ex4}
L(z) = \begin{pmatrix}
zP'_1 & zR_1Q_2 & zR_1P_2Q_3 & zR_1P_{2,3} \\
0     & zP'_2   & zR_2Q_3    & zR_2P_3     \\
0     & 0       & zP'_3      & zR_3 \\
Q_1   & P_1Q_2  & P_{1,2}Q_3 & P_{1,3}
\end{pmatrix}.
\end{equation}
Our convention is 
$L(z)(\alpha \ot v_j) = 
\sum_i(L_{ij}(z)\alpha)\ot v_i$ for $\alpha  \in {\mathcal A}$.
Similarly we let 
$\overset{1}{L}(z) , \overset{2}{L}(z) \in 
{\mathcal A} \ot {\rm End}(V \ot V)$ 
denote the operators acting as 
$\overset{1}{L}(z)(\alpha \ot v_i \ot v_j) 
= \sum_k (L_{ki}(z)\alpha)\ot v_k \ot v_j$ and 
$\overset{2}{L}(z)(\alpha \ot v_i \ot v_j) 
= \sum_k (L_{kj}(z)\alpha)\ot v_i \ot v_k$.
As an analogue of the Yang-Baxter equation (\ref{eqa:ybe2}), we have
\begin{proposition}\label{pra:RLL}
\begin{equation*}
R(z_2/z_1)\overset{2}{L}(z_2)\overset{1}{L}(z_1) 
= \overset{1}{L}(z_1)\overset{2}{L}(z_2)R(z_2/z_1) \in 
{\mathcal A}\ot{\rm End}(V \ot V).
\end{equation*}
\end{proposition}
In section \ref{subsec:facL}, this will be proved based on 
the factorization of $L(z)$.

\subsection{\mathversion{bold}
Factorization of $L(z)$}\label{subsec:facL}
Let us introduce the operators 
$K_i \in {\mathcal A} \ot  {\rm End }(V)$ for $1 \le i \le n-1$ by
\begin{equation}\label{eqa:kdef}
\begin{split}
&K_i = ((K_{i})_{j,k})_{1 \le j,k \le n},\\
&(K_{i})_{i,i} = P'_i,\;
(K_{i})_{i,n} = R_i,\;
(K_{i})_{n,i} = Q_i,\;
(K_{i})_{n,n} = P_i,\\
&(K_{i})_{j,j} = 1\,(j \neq i, n).
\end{split}
\end{equation}
The other elements are zero.
The $K_i$ with $\forall a_i=1$ will be interpreted as the 
local propagation 
operator in quantized box-ball system in section \ref{subsec:qbbs3}.
We also introduce an $n$ by $n$ matrix 
\begin{equation*}
D(z) = z \hbox{ diag}(1,\ldots, 1,z^{-1}),
\end{equation*}
which acts on $V$ only.
\begin{proposition}\label{pra:factor}
\begin{equation*}
L(z) = D(z)K_1K_2 \cdots K_{n-1}
\end{equation*}
\end{proposition}
For example the latter in (\ref{eqa:ex23}) is expressed as
\begin{equation*}
\begin{pmatrix}
zP'_1 & zR_1Q_2 & zR_1P_2\\
0    & zP'_2   & zR_2\\
Q_1 & P_1Q_2& P_1P_2
\end{pmatrix} = \text{diag}(z,z,1)
\begin{pmatrix}
P'_1 & 0 & R_1\\
0    & 1   & 0\\
Q_1 & 0 & P_1
\end{pmatrix}
\begin{pmatrix}
1 & 0 & 0 \\
0    & P'_2   & R_2\\
0 & Q_2& P_2
\end{pmatrix}.
\end{equation*}
\begin{proof}
Denote the $n$ by $n$ matrix $L(z\!=\!1)$ defined by (\ref{eqa:L-elements}) 
by $L_n$.
We are to show $K_1K_2\ldots K_{n-1} = L_n$ for $A^{(1)}_{n-1}$. 
This is done by induction on $n$. The case $n=3$ is 
checked in the above.
Suppose the equality is valid for $n$.
Then from the structure of the matrices $K_i$, 
one can evaluate $K_1K_2 \cdots K_{n}$ for $A^{(1)}_n$ 
as the product of $K_1$ and the rest as
\begin{equation}\label{eqa:kpro}
\begin{small} \begin{pmatrix}
P'_1 & & R_1 \\
& & \\
    & \openone_{n-1} & \\
& &         \\
Q_1 & & P_1 
\end{pmatrix}
\end{small}
\begin{pmatrix}
1 & 0 & \cdots & 0\\
0 &  &  & \\
\vdots &  & L_n^+  \\
0 &  & &
\end{pmatrix} = L_{n+1}.
\end{equation}
Here $L_n^+$ is $L_n$ with all the constituent operators 
$X_i (X = P, P', Q, R)$ replaced by $X_{i+1}$, and 
$\openone_{n-1}$ is the identity matrix of size $n-1$.
It is straightforward to verify this identity.
\end{proof}
\begin{remark}\label{rema:com}
Elements of ${\mathcal A}$ contained in any single $L_{ij}(z)$ 
(\ref{eqa:L-elements}) are all commutative.
As a result, the identity (\ref{eqa:kpro}) holds 
under any interchange of 
$P_i, P'_i, Q_i$ and $R_i$ on the both sides.
\end{remark}
Let us make use of the factorization to prove 
Proposition \ref{pra:RLL}.
We first define $\sigma_1, \ldots, \sigma_{n-1}, 
\sigma \in {\rm End }(V)$ by 
\begin{align*}
&\sigma_iv_j = \begin{cases}
v_{i+1} & \hbox{ if } j=i\\
v_i & \hbox{ if } j=i+1\\
v_j & \hbox{ otherwise,} 
\end{cases}\\
&\sigma = \sigma_{n-1}\sigma_{n-2} \cdots \sigma_1.
\end{align*}
Thus $\sigma v_j = v_{j-1}$ is valid for indices in $\Z/n\Z$.
Consider the following gauge transformation of $K_i$:
\begin{equation}\label{eqa:sdef1}
S_i = \sigma_i \sigma_{i+1}\cdots\sigma_{n-1}K_i
\sigma_{n-1}\sigma_{n-2}\cdots \sigma_{i+1} \quad 1 \le i \le n-1.
\end{equation}
The components of $S_i \in {\mathcal A} \ot {\rm End }(V)$ 
are given by
\begin{equation}\label{eqa:sdef2}
\begin{split}
&S_i = ((S_{i})_{j,k})_{1 \le j,k \le n},\\
&(S_{i})_{i,i+1} = P_i,\;
(S_{i})_{i+1,i+1} = R_i,\;
(S_{i})_{i,i} = Q_i,\;
(S_{i})_{i+1,i} = P'_i,\\
&(S_{i})_{j,j} = 1\,(j \neq i, i+1).
\end{split}
\end{equation}
The other components are zero.
Note that Proposition \ref{pra:factor} is rewritten as
\begin{equation}\label{eqa:ls}
L(z) = D(z)\sigma S_1 S_2 \cdots S_{n-1}.
\end{equation}
Now Proposition \ref{pra:RLL} is a corollary of
the formula (\ref{eqa:ls}) and 
\begin{lemma}\label{lema:rss}
\begin{align*}
&R(z_2/z_1)(D(z_1)\sigma \ot D(z_2)\sigma)
= (D(z_1)\sigma \ot D(z_2)\sigma)R(z_2/z_1),\\
&R(z)\overset{2}{S}_i\overset{1}{S}_i 
= \overset{1}{S}_i\overset{2}{S}_iR(z)\quad 1 \le i \le n-1.
\end{align*}
\end{lemma}
\begin{proof}
The first relation is directly confirmed.
It is enough to check the latter at two distinct values of $z$.
It is trivially valid at $z=1$ and easily checked at 
$z=q^{-2}$.
\end{proof}
\begin{remark}\label{rema:K2}
If $a_i = 1$,  the property
\begin{equation*}
K^2_i = \openone_n
\end{equation*}
is valid for $1 \le i \le n-1$. This is a remnant of 
the inversion relation (\ref{eqa:inv}).
It implies $L(z)^{-1} = K_{n-1}\cdots K_1D(z^{-1})$.
The formula (\ref{eqa:ls}) was known 
at $q=0$ as a factorization of combinatorial $R$ 
\cite{HKT2}, where $S_i$ appeared as the Weyl group operator on 
crystal basis.
\end{remark}
For $A^{(1)}_1$, the $L$ operator here can also be obtained by
specializing the $q$ generic case of the one in \cite{BS}.
The case $\forall a_i = 0$
has appeared in the 
quantized Volterra model for $A^{(1)}_{n-1}$ \cite{HIK}.

\subsection{Quantized box-ball system: 
Space of states}\label{subsec:qbbs1}

Consider the formal infinite tensor product 
of $V = \C v_1 \oplus \cdots \oplus \C v_n$:
\begin{equation}\label{eqa:VV}
\cdots \ot V \ot V \ot V \ot \cdots = \oplus\; \C \,
(\cdots \ot v_{j_{-1}}\ot v_{j_0} \ot v_{j_1} \ot \cdots).
\end{equation} 
An element of the form 
$c(\cdots \ot v_{j_{-1}}\ot v_{j_0} \ot v_{j_1} \ot \cdots)$ 
will be 
called a monomial (a monic monomial if with $c=1$). 
The space of states of our quantized box-ball system is 
the subspace of (\ref{eqa:VV}) given by
\begin{equation}\label{eqa:pdef}
{\mathcal P} = \{ \sum_{p: \text{monic monomial}} c_p p  
\mid  \hbox{conditions (i) and (ii)} \},
\end{equation}
where 
(i) $\sum_{k \in \Z} \vert j_k -n \vert < \infty$ 
for any $p = \cdots \ot v_{j_{-1}}\ot v_{j_0} \ot v_{j_1} \ot \cdots$ 
appearing in the sum,
(ii) there exists $N \in \Z$ such that 
$\lim_{q \rightarrow 0} q^N\sum_p c_p p = 0$.
Monomials can be classified according to 
the numbers $w_1,\ldots, w_{n-1}$ of occurrence of the 
letters $1, \ldots, n-1$ in the set $\{j_k\}$.
Consequently one has the direct sum decomposition:
\begin{equation}\label{eqa:dsd}
{\mathcal P} = \oplus {\mathcal P}_{w_1,w_2,\ldots, w_{n-1}},
\end{equation}
where the sum runs over $(w_1,\ldots, w_{n-1}) \in \Z^{n-1}_{\ge 0}$.
We have ${\mathcal P}_{0,\ldots,0}= \C p_{\rm vac}$, where
$p_{\rm vac} = \cdots \ot v_n \ot v_n \ot \cdots$.

The local states $v_{j_k} \in V$ is regarded as 
the $k$th box containing a ball with color $j_k$ if 
$j_k \neq n$, and the empty box if $j_k=n$.
The space of states of the box-ball system is the 
totality of the monomials in the above sense.
The space of states ${\mathcal P}$ of our
quantized box-ball system consists of 
linear superpositions thereof.

\subsection{Time evolution}\label{subsec:qbbs2}

We set $\forall a_i = 1$ in the remainder of section \ref{sec:A}.
Then the following provides an 
${\mathcal A}$ module ${\mathcal M}$:
\begin{equation}\label{eqa:M}
\begin{split}
&{\mathcal M} = \oplus_{m_1, \ldots, m_{n-1} \in \Z_{\ge 0}}
\C [m_1,\ldots, m_{n-1}],\\
&P_i[\ldots, m_i,\ldots] = q^{m_i}[\ldots, m_i,\ldots],\\
&Q_i[\ldots, m_i,\ldots] = [\ldots,m_i\!+\!1,\ldots],\\
&R_i[\ldots, m_i,\ldots] = (1-q^{2m_i})[\ldots, m_i\!-\!1,\ldots],
\end{split}
\end{equation}
where the right hand side of the last formula 
is to be understood as 0 at $m_i=0$.
The space ${\mathcal M}$ will be regarded as the space of the 
quantum carrier.
By construction, for $x = [x_1,\ldots, x_{n-1}] \in {\mathcal M}$ one has
\begin{equation}\label{eqa:LW}
\begin{split}
&L(z)(x \ot v_j) = \sum_kW_{jk}[x \vert y](y \ot v_k),\\
&W_{jk}[x \vert y]= \lim_{x_n \rightarrow \infty}
w_{jk}[x_1,\ldots,x_{n-1},x_n \vert y_1,\ldots,y_{n-1},y_n],
\end{split}
\end{equation}
where $y$ is determined {}from (\ref{eqa:y}) 
in terms of $j,k$ and $x$.

According to the standard construction of transfer matrices in 
two dimensional solvable vertex models \cite{Bax}, the time evolution 
$T(z): {\mathcal P} \rightarrow {\mathcal P}$ is
constructed as a composition of 
local $L$ operators as
\begin{equation}\label{eqa:defT}
T(z) = 
\bigl(\cdots \overset{1}{L}(z)
\overset{0}{L}(z)
\overset{-1}{L}(z) \cdots \bigr)_{0,0}.
\end{equation}
Here $\overset{k}{L}(z) 
\in {\rm End }({\mathcal M} \ot {\mathcal P})$ signifies 
the representation of the $L$ operator:
\begin{equation}
\begin{split}
&\overset{k}{L}(z)\bigl(m \ot 
(\cdots \ot v_{j_{k-1}} \ot v_{j_k} \ot v_{j_{k+1}} \ot \cdots) \bigr)\\
&=\sum_i \bigl(L_{ij_k}(z)m\bigr) \ot 
(\cdots \ot v_{j_{k-1}} \ot v_{i} \ot v_{j_{k+1}} \ot \cdots),
\end{split}
\end{equation}
where $L_{ij_k}(z)m$ for $m \in {\mathcal M}$ is specified by 
(\ref{eqa:M}).
The symbol  $(\cdots)_{0,0}$ in (\ref{eqa:defT}) stands for
the element in ${\rm End }({\mathcal P})$ 
that is attached to the transition  
$[0,\ldots,0] \mapsto [0,\ldots,0]$ in the 
${\mathcal M}$ part.
By the definition 
$T(z)$ preserves the 
weight subspace ${\mathcal P}_{w_1,w_2,\ldots, w_{n-1}}$ and 
acts homogeneously on it as
\begin{equation}\label{eqa:homo}
T(z)p = z^{w_1+\cdots+w_{n-1}}T(1)p\quad \hbox{for } 
p \in {\mathcal P}_{w_1,w_2,\ldots, w_{n-1}}.
\end{equation}
Therefore the commutativity $T(z)T(z') = T(z')T(z)$ is trivially valid.
Henceforth we concentrate on $T=T(z\!=\!1)$, and 
$T(p)$ for $p \in {\mathcal P}$ is to be understood as $T(1)p$.

\subsection{Factorized dynamics}\label{subsec:qbbs3}

The time evolution $T$ admits a simple description 
as the product of propagation operators.
Set
\begin{equation}\label{eqa:pKdef1}
{\mathcal K}_i 
= \bigl(\cdots \overset{1}{K}_{i}
\overset{0}{K}_{i}
\overset{-1}{K}_{i} \cdots \bigr)_{0,0} 
\in {\rm End }({\mathcal P}) \quad 
1 \le i \le n-1,
\end{equation}
where the representation $\overset{k}{K}_{i} \in 
{\rm End }({\mathcal M}\ot{\mathcal P})$ is specified {}from
$K_i$ (\ref{eqa:kdef}) in the same way as
$\overset{k}{L}(z)$ was done via $L(z)$.
To interpret ${\mathcal K}_i$ pictorially, 
we attach the following diagrams 
to the local operator $K_i$.

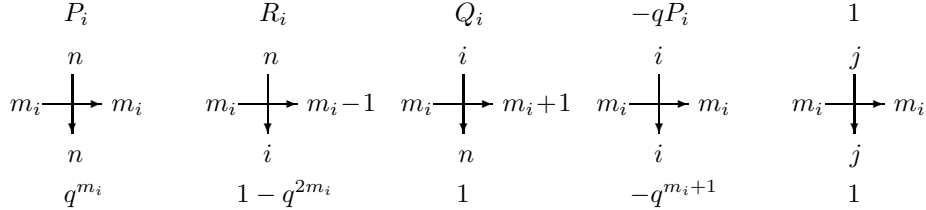
\begin{figure}[h]
\setlength{\unitlength}{1.3mm}
\caption{Diagram for $K_i\; (j \neq i,n)$}\label{fig:Ki}
\begin{picture}(125,23)(0,-0.7)
\multiput(0,0)(20,0){5}{
\put(5,8){\vector(1,0){6}}\put(8,11){\vector(0,-1){6}}
\put(1.6,7.3){$m_i$}}

\put(12,7.3){$m_i$}
\put(32,7.3){$m_i\!-\!1$}
\put(52,7.3){$m_i\!+\!1$}
\put(72,7.3){$m_i$}
\put(92,7.3){$m_i$}

\put(7.5,12.1){$n$}\put(7.5,2.1){$n$}
\put(27.5,12.1){$n$}\put(27.5,2.1){$i$}
\put(47.5,12.1){$i$}\put(47.5,2.1){$n$}
\put(67.5,12.1){$i$}\put(67.5,2.1){$i$}
\put(87.5,12.1){$j$}\put(87.5,2.1){$j$}

\put(7.1,-2){$q^{m_i}$}\put(7.1,16.5){$P_i$}
\put(24.8,-2){$1-q^{2m_i}$}\put(27.1,16.5){$R_i$}
\put(47.3,-2){$1$}\put(47.1,16.5){$Q_i$}
\put(65,-2){$-q^{m_i+1}$}\put(65.1,16.5){$-qP_i$}
\put(87.3,-2){$1$}\put(87.3,16.5){$1$}

\end{picture}
\end{figure}

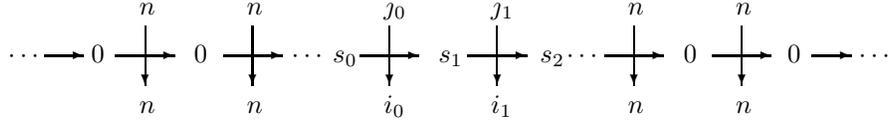
\begin{figure}[h]
\setlength{\unitlength}{1.3mm}
\caption{Diagram for ${\mathcal K}_i\; (j \neq i,n)$}
\label{fig:calKi}
\begin{picture}(90,16)(-5,3)
\multiput(0,0)(25,0){3}
{
\put(5,8){\vector(1,0){6}}\put(8,11){\vector(0,-1){6}}
\put(16,8){\vector(1,0){6}}\put(19,11){\vector(0,-1){6}}
}

\put(-2.3,8){\vector(1,0){4}}\put(76.2,8){\vector(1,0){4}}

\put(2.5,7.3){$0$}\put(13,7.3){$0$}
\put(27.2,7.3){$s_0$}\put(38,7.3){$s_1$}\put(48.4,7.3){$s_2$}
\put(63.1,7.3){$0$}\put(73.7,7.3){$0$}

\put(-6,7.3){$\cdots$}\put(23,7.3){$\cdots$}
\put(51.1,7.3){$\cdots$}\put(81,7.3){$\cdots$}

\put(7.4,12.2){$n$}\put(18.4,12.2){$n$}
\put(32.4,12.2){$j_0$}\put(43.4,12.2){$j_1$}
\put(57.4,12.2){$n$}\put(68.4,12.2){$n$}

\put(7.4,2.2){$n$}\put(18.4,2.2){$n$}
\put(32.4,2.2){$i_0$}\put(43.4,2.2){$i_1$}
\put(57.4,2.2){$n$}\put(68.4,2.2){$n$}

\end{picture}
\end{figure}

\noindent
Here $m_i \in \Z_{\ge 0}$ is a coordinate in 
$[m_1,\ldots, m_{n-1}] \in {\mathcal M}$.
The horizontal and vertical arrows correspond to 
${\mathcal M}$ and $V$, respectively.
The diagrams depict the interaction between 
the local box and 
the quantum carrier containing $m_i$ balls of color $i$.
The carrier coming {}from the left encounters the local box
whose state are specified on the top.
It picks up/down a color $i$ ball or does nothing 
and proceeds 
to the right leaving the box in the state given in the bottom  
with the listed amplitudes.
The first line in the figure gives the operators acting on 
${\mathcal M}$ that yield the amplitudes on the last line.
For example one has 
\begin{align*}
K_i([\ldots,m_i,\ldots]\ot v_n) 
&= (P_i[\ldots,m_i,\ldots]) \ot v_n +
(R_i[\ldots,m_i,\ldots]) \ot v_i\\
&= q^{m_i}[\ldots,m_i,\ldots] \ot v_n + 
(1-q^{2m_i})[\ldots, m_i-1,\ldots] \ot v_i.
\end{align*}
The second term describes unloading whereas the first 
term is just a passage.  
It is easy to see that at $q=0$, $K_i$ reduces to the  
deterministic operator which coincides with the 
local interaction between a carrier and a box \cite{TM} in the 
conventional box-ball system \cite{T,TS}.
Now the composition (\ref{eqa:pKdef1}) 
is expressed as Fig. \ref{fig:calKi}.

\noindent
The amplitude of ${\mathcal K}_i$ 
assigned with the transition {}from 
$(\cdots \ot v_{j_0}\ot v_{j_1} \ot \cdots)$ to 
$(\cdots \ot v_{i_0}\ot v_{i_1} \ot \cdots)$ is obtained
as the product of all the amplitudes attached to 
the local vertices in Fig. \ref{fig:calKi} according to the 
rule specified in Fig. \ref{fig:Ki}.
The calculation involves 
an infinite product, which is well defined for elements in 
${\mathcal P}$.
See section \ref{subsec:norm}
for examples of computations of the amplitudes.
\begin{theorem}\label{th:facT}
The time evolution of the quantized box-ball system admits a
factorization into propagation operators as
\begin{equation}
T = {\mathcal K}_1 \cdots {\mathcal K}_{n-1}.
\end{equation}
\end{theorem}
\begin{proof}
This is a consequence of the definitions 
(\ref{eqa:defT}), (\ref{eqa:pKdef1}) and 
the factorization of the $L$ operator established in 
Proposition \ref{pra:factor}.
\end{proof}
At $q=0$, Theorem \ref{th:facT} reduces to the 
original description of the time evolution in 
the box-ball system \cite{T} as the composition of 
finer process to move balls with a fixed color.

\subsection{Some properties of amplitudes}
\label{subsec:norm}

For simplicity we concentrate on $A^{(1)}_1$ case 
in the remainder of section \ref{sec:A}, where one only has one kind of ball 
and $T= {\mathcal K}_1$. 
However, by virtue of Theorem \ref{th:facT}, 
all the essential statements are equally  
valid for general $A^{(1)}_{n-1}$ 
under an appropriate resetting.
In particular, Proposition \ref{pra:TT} and 
Proposition \ref{pra:norm1} remain valid not only for $T$ but also 
${\mathcal K}_i$ for any $1 \le i \le n-1$.

Let us write the action of the time evolution of 
a monic monomial 
$p \in {\mathcal P}$ as 
$T(p) = \sum_{p'}A_{p',p}p'$, where the sum is taken over 
monic monomials $p' \in {\mathcal P}$.
We then define the transposition ${}^tT$ of $T$ by
${}^tT(p) = \sum_{p'}A_{p,p'}p'$.
\begin{proposition}\label{pra:TT}
\begin{equation}
{}^tT = T^{-1}
\end{equation}
\end{proposition}
\begin{proof}
In view of Remark \ref{rema:K2}, the inverse 
$T^{-1}= {\mathcal K}^{-1}_1$ is obtained by reversing the 
horizontal arrows in Fig. \ref{fig:Ki} and 
sending the carrier {}from the right to the left 
correspondingly in Fig. \ref{fig:calKi}.
By using this fact, one can verify the claim.
See also Remark \ref{rema:inv}.
\end{proof}
Let $(\;,\;)$ be the inner product 
such that $(p,p') = \delta_{p,p'}$ for all the 
monic monomials $p$ and $p'$.
It is well defined on a subset of ${\mathcal P} \times {\mathcal P}$.
Then Proposition \ref{pra:TT} tells that 
$(T(r),T(s)) = (r,s)$ for $(r,s)$ belonging to the subset.
This property leads to a family of $q$-series identities.
In fact one has $\sum_p A_{p,r}A_{p,s} = \delta_{r s}$ for 
any monic monomials $r$ and $s$.
Pick the monomial 
$p = \cdots \ot v_2 \ot v_1 \ot v_2 \ot \cdots$ for instance.
Then the left hand side of $(T(p),T(p))=1$, 
the sum of squared amplitudes, is calculated as 
\begin{equation*}
(-q)^2 + \sum_{k \ge 0}\bigl(q^k(1-q^2)\bigr)^2 = 1.
\end{equation*}
Similarly for the monomial 
$p = \cdots \ot v_2 \ot v_1 \ot v_1 \ot v_2 \ot \cdots$, 
the contributions to $(T(p),T(p))=1$ are grouped into 
the four cases as in Fig. \ref{fig:2sol}, which add up to 1.

\begin{figure}[h]
\setlength{\unitlength}{1.3mm}
\caption{Squared amplitudes for $T(p)$}\label{fig:2sol}
\begin{picture}(50,42)(25,0)
\multiput(0,0)(0,10){4}
{\put(5,8){\circle*{1}}\put(8,8){\circle*{1}}}

\put(5,35){\circle*{1}}\put(8,35){\circle*{1}}
\put(30,36){$(-q)^4$}

\put(5,25){\circle*{1}}\put(8,25){\circle{1}}
\put(9,25){$\underbrace{\cdots}_{k}$}
\put(15,25){\circle*{1}}
\put(30,26){$(-q)^2\sum_{k \ge 0}
(q^k)^2(1-q^2)^2 = q^2(1-q^2)$}

\put(5,15){\circle{1}}\put(8,15){\circle*{1}}
\put(9,15){$\underbrace{\cdots}_{k}$}
\put(15,15){\circle*{1}}
\put(30,16){$(-q^2)^2\sum_{k \ge 0}
(q^k)^2(1-q^2)^2 = q^4(1-q^2)$}

\put(5,5){\circle{1}}\put(8,5){\circle{1}}
\put(9,5){$\underbrace{\cdots}_{k_1}$}
\put(15,5){\circle*{1}}
\put(16.5,5){$\underbrace{\cdots}_{k_2}$}
\put(22.5,5){\circle*{1}}
\put(30,6){$\sum_{k_1,k_2 \ge 0}
(q^{2k_1+k_2})^2(1-q^2)^2(1-q^4)^2 = 
(1-q^2)(1-q^4)$}

\end{picture}
\end{figure}

\noindent
Here the symbols $\bullet$ and  
$\circ$ stand for a ball $v_1$ and an empty box $v_2$, respectively.
The symbol $\cdots$ represents an array of empty boxes 
of the specified number.
In each group, the upper configuration is $p$ and the 
lower one is a monomial occurring in $T(p)$. 

So far we have considered the 
quadratic form $(\;,\;)$.
Now we turn to a linear one.
We use the standard notation
\begin{align*}
&(z)_m = (z;q)_m = (1-z)(1-zq)\cdots (1-zq^{m-1}),\\
&\left[ \begin{array}{c} m  \\ k \end{array} \right] = 
\frac{(q)_m}{(q)_{k}(q)_{m-k}}.
\end{align*}

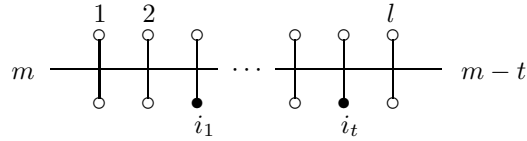
\begin{figure}[h]
\setlength{\unitlength}{1.3mm}
\caption{$\beta_{m,t,l}$}\label{fig:beta}
\begin{picture}(100,16)(-20,2)
\put(14.4,12.8){$1$}\put(19.4,12.8){$2$}\put(44.4,12.8){$l$}
\put(10,8){\line(1,0){17}}\put(28.5,7.3){$\cdots$}\put(33,8){\line(1,0){17}}
\multiput(15,5)(5,0){3}{\put(0,0){\line(0,1){6}}}
\multiput(35,5)(5,0){3}{\put(0,0){\line(0,1){6}}}
\put(6,7){$m$}\put(52,7){$m-t$}
\multiput(15,11.5)(5,0){3}{\put(0,0){\circle{1}}}
\multiput(35,11.5)(5,0){3}{\put(0,0){\circle{1}}}
\put(25,4.5){\circle*{1}}\put(40,4.5){\circle*{1}}
\put(15,4.5){\circle{1}}\put(20,4.5){\circle{1}}
\put(35,4.5){\circle{1}}\put(45,4.5){\circle{1}}
\put(24.7,1.5){$i_1$}\put(39.7,1.5){$i_t$}
\end{picture}
\end{figure}

For $t \le \min(l,m)$, 
let $\beta_{m,t,l}$ be the sum of all the amplitudes for 
$l$ successive vacant boxes to acquire $t$ balls 
during the passage of a carrier containing $m$ balls.
Namely, it is the sum of the amplitudes 
for Fig. \ref{fig:beta} over $1 \le i_1 < i_2 < \cdots < i_t \le l$.

\begin{lemma}\label{lema:beta}
\begin{equation}
\beta_{m,t,l} = q^{(m-t)(l-t)}
(1-q^{2m})(1-q^{2m-2})\cdots (1-q^{2(m-t+1)})
\left[ \begin{array}{c} l  \\ t \end{array} \right].
\end{equation}
\end{lemma}
\begin{proof}
The contribution {}from Fig. \ref{fig:beta} is 
\begin{equation}
\begin{split}
&(1-q^{2m})(1-q^{2m-2})\cdots (1-q^{2(m-t+1)})\\
&\times
q^{m(i_1-1)+(m-1)(i_2-i_1-1)+\cdots + (m-t+1)(i_t-i_{t-1}-1)
+(m-t)(l-i_t)}.
\end{split}
\end{equation}
The claim follows by summing this over 
$1 \le i_1 < i_2 < \cdots < i_t \le l$.
\end{proof}

Let ${\mathcal P}_{\rm fin}$ be the subspace of 
${\mathcal P}$ spanned by the superpositions of 
monomials $\sum_p c_p p$ in which $\sum_p c_p$ exists.
For instance, monomials are elements of ${\mathcal P}_{\rm fin}$.
Consider the linear function 
${\mathcal N}: {\mathcal P}_{\rm fin} \rightarrow \C$
that takes value $1$ on all the monic monomials.
\begin{proposition}\label{pra:norm1}
$T$ preserves ${\mathcal N}$, i.e., 
${\mathcal N}(T(p)) = {\mathcal N}(p)$ 
for any $p \in {\mathcal P}_{\rm fin}$.
\end{proposition}

For example for $p = \cdots v_2 \ot v_1 \ot v_2 \ot \cdots$, one has
\begin{equation*}
{\mathcal N}(p) = -q + \sum_{k \ge 0}q^k(1-q^2) = 1.
\end{equation*}
For $p = \cdots v_2 \ot v_1 \ot v_1 \ot v_2 \ot \cdots$
considered in Fig. \ref{fig:2sol}, one has
\begin{equation*}
{\mathcal N}(p) = (-q)^2 - q\sum_{k \ge 0}q^k(1-q^2)
- q^2\sum_{k \ge 0}q^k(1-q^2) + 
\sum_{k_1,k_2 \ge 0}q^{2k_1+k_2}(1-q^2)(1-q^4) = 1.
\end{equation*}

The remainder of this section \ref{subsec:norm} is devoted to a proof of 
Proposition \ref{pra:norm1}.
We begin by introducing a map $\Phi_m$ for $m \in \Z_{\ge 0}$, which is 
a slight generalization of $T$.
We set $\Phi_0 = T$. 
For $m \ge 1$, $\Phi_m$ acts on 
${\mathcal P}_{\rm fin}\setminus\{p_{\rm vac}\}$ as follows.
Pick any monic monomial 
$p \in {\mathcal P}_{\rm fin}\setminus\{p_{\rm vac}\}$ 
and decompose it uniquely as
$p = p_{{\rm {\small left}}} \ot p_{{\rm {\small right}}}$,  
so that $p_{{\rm {\small left}}}$ is free of balls and 
the leftmost component of $p_{{\rm {\small right}}}$ is a ball.
Let $p'_{{\rm {\small right}}}$ be the linear combination of 
the monic monomials generated by the penetration 
of the carrier initially containing
$m$ balls through $p_{{\rm {\small right}}}$ to the right.
See Fig. \ref{fig:p'}.

\begin{figure}[h]
\setlength{\unitlength}{1.3mm}
\caption{$p'_{{\rm {right}}}$}\label{fig:p'}
\begin{picture}(60,20)(-7,-2)
\put(14.4,12.8){$\overbrace{\qquad\qquad\qquad}^{p_{{\rm {\small right}}}}$}
\put(10,8){\line(1,0){17}}
\multiput(15,5)(5,0){2}{\put(0,0){\line(0,1){6}}}

\put(6,7){$m$}\put(29,7){$\cdots$}
\put(15,11.5){\circle*{1}}
\put(14.4,4){$\underbrace{\qquad\qquad\qquad}_{p'_{{\rm {\small right}}}}$}
\end{picture}
\end{figure}

We set 
$\Phi_m(p) = p_{{\rm {\small left}}} \ot p'_{{\rm {\small right}}}$
and extend it linearly to the map 
$\Phi_m: {\mathcal P}_{\rm fin}
\setminus\{p_{\rm vac}\} \rightarrow 
{\mathcal P}_{\rm fin}\setminus\{p_{\rm  vac}\}$.
It is a direct sum of the action 
${\mathcal P}_{{\rm fin},N} 
\rightarrow {\mathcal P}_{{\rm fin},N+m}$ over $N \in \Z_{\ge 1}$,
where the notation 
${\mathcal P}_{{\rm fin},N}$ is the $n=2$ case of (\ref{eqa:dsd}) 
restricted to ${\mathcal P}_{\rm fin}$.

Proposition \ref{pra:norm1} is obvious for $p = p_{\rm vac}$.
Since ${\mathcal N}$ is linear, the other case follows {}from 
the $m=0$ case of 
\begin{proposition}\label{pra:alpha}
For any monic monomial $p \in {\mathcal P}_N$,
\begin{equation}\label{eqa:alpha}
{\mathcal N}(\Phi_m(p)) = (1+q)(1+q^2)\cdots (1+q^m)
\end{equation}
is valid for any $m \ge 0$ and  $N \ge 1$.
\end{proposition}
The right hand side depends on $m$ but not on $N$, hence it will be 
denoted by $\alpha_m$.  Note that $\alpha_m = \beta_{m,m,\infty}$.
\begin{proof}
We show (\ref{eqa:alpha}) by induction on $N$. 
For $N=1$, the relevant configurations either accommodate a ball or not 
just below the initial one. The former contributes 
$-q^{m+1}\beta_{m,m,\infty}$ to ${\mathcal N}(\Phi_m(p))$ and the latter does 
$\beta_{m+1,m+1,\infty}$.  The two contributions indeed sum up to $\alpha_m$.
Assume the claim for $N$.
In the monic monomial $p \in {\mathcal P}_{N+1}$, suppose 
there are $l$ empty boxes between the leftmost ball 
and its nearest neighbor. 
The configurations that accommodate $t$ balls in the 
$l$ boxes are classified into the two cases in Fig. \ref{fig:two}.

\begin{figure}[h]
\setlength{\unitlength}{1.3mm}
\caption{Two kinds of contributions to 
${\mathcal N}(\Phi_m(p))$}\label{fig:two}
\begin{picture}(100,32)(-20,-17)
\multiput(0,0)(0,-17){2}
{\put(10,8){\line(1,0){13}}\put(51,8){\line(1,0){8}}
\put(25,7.3){$\cdots$}\put(61.5,7.3){$\cdots$}
\put(30,8){\line(1,0){10}}

\put(15,5){\line(0,1){6}}
\put(20,5){\line(0,1){6}}
\put(35,5){\line(0,1){6}}
\put(55,5){\line(0,1){6}}
\put(6,7){$m$}

\put(15,11.5){\circle*{1}}
\put(20,11.5){\circle{1}}
\put(35,11.5){\circle{1}}
\put(55,11.5){\circle*{1}}
\put(20,4){$\underbrace{\qquad\qquad\qquad}_{t \;{\rm balls}}$}}

\put(43,7){$m\!-\!t$}
\put(41,-10){$m\!+\!1\!-\!t$}

\put(15,4.5){\circle*{1}}
\put(15,-12.5){\circle{1}}

\end{picture}
\end{figure}
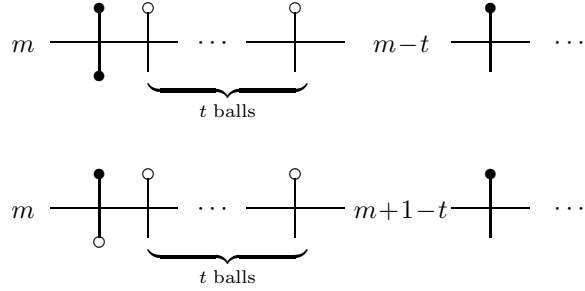
Accordingly we have the recursion relation
\begin{equation}
{\mathcal N}(\Phi_m(p)) = -q^{m+1}
\sum_{t=0}^{\min(l,m)}\beta_{m,t,l}{\mathcal N}(\Phi_{m-t}({\tilde p}))+
\sum_{t=0}^{\min(l,m+1)}\beta_{m+1,t,l}
{\mathcal N}(\Phi_{m+1-t}({\tilde p})),
\end{equation}
where ${\tilde p} \in {\mathcal P}_N$ is the monic monomial obtained 
by removing the leftmost ball {}from $p$.
Thus we are done if 
\begin{equation*}
\alpha_m = -q^{m+1}
\sum_{t=0}^{\min(l,m)}\beta_{m,t,l}\alpha_{m-t}+
\sum_{t=0}^{\min(l,m+1)}\beta_{m+1,t,l}\alpha_{m+1-t}
\end{equation*}
is shown. This is a corollary of Lemma \ref{lema:qpol}.
\end{proof}

\begin{lemma}\label{lema:qpol}
Let $l,m \in \Z_{\ge 0}$. Then 
\begin{equation*}
\alpha_m = \sum_{t=0}^{\min(l,m)}\beta_{m,t,l}\alpha_{m-t}.
\end{equation*}
\end{lemma}
\begin{proof}
We are to show
\begin{equation*}
1 = \sum_{t=0}^{\min(l,m)}q^{(l-t)(m-t)}
\frac{(q)_l(q)_m}{(q)_t(q)_{l-t}(q)_{m-t}}.
\end{equation*}
Since the both sides are symmetric with respect to $l$ and $m$,
we assume with no loss of generality that $l\le m$.
Applying the $q-$binomial identity 
$(z;q)_t = \sum_{s=0}^t
\left[ \begin{array}{c} t  \\ s \end{array} \right]
(-z)^sq^{s(s-1)/2}$, we expand the factor 
$(q)_m/(q)_{m-t} = (q^{m-t+1};q)_t$.
Then the right hand side becomes
\begin{equation*}
\sum_{t=0}^l\sum_{s=0}^t(-1)^sq^{(m-t)(l-t+s)+s(s+1)/2}
\frac{(q)_l}{(q)_{l-t}(q)_s(q)_{t-s}}.
\end{equation*}
By eliminating $t$ by setting $t=s+i$, this is written as 
\begin{equation*}
\sum_{i=0}^l
\left[ \begin{array}{c} l  \\ i \end{array} \right]q^{(m-i)(l-i)}
\sum_{s=0}^{l-i}
\left[ \begin{array}{c} l-i  \\ s \end{array} \right]
(-q^{i-l+1})^sq^{s(s-1)/2}.
\end{equation*}
The $q-$binomial identity tells that the sum over $s$ is 
equal to $(q^{i-l+1};q)_{l-i}=\delta_{il}$.
\end{proof}
\begin{remark}\label{rema:eigen}
Set $u = \sum_p p$, where the sum extends over all the 
monic monomials in ${\mathcal P}_N$ for any $N \ge 0$.
Then Proposition \ref{pra:TT} and Proposition \ref{pra:norm1} 
tell that $T(u) = u$. Conversely, this property and 
Proposition \ref{pra:TT} imply Proposition \ref{pra:norm1} 
since ${\mathcal N}(p) = (p,u) = (T(p),T(u)) = (T(p),u) = 
{\mathcal N}(T(p))$.
\end{remark}

\subsection{Bethe ansatz}\label{subsec:ba}
Consider the commuting family of 
transfer matrices $T_m(z)\; (m \in \Z_{\ge 1})$ constructed 
{}from the fusion $R$ matrix $R^{(m,1)}(z)$ (\ref{eqa:Rm1w}).
Normalize them so that $T_m(z)p_{\rm vac} = p_{\rm vac}$.
Then the time evolution $T$ of our quantized box-ball system 
belongs to the family as $T=T_\infty(1)$.
It therefore shares the eigenvectors with the simplest one $T_1(z)$, 
which corresponds to the well known six vertex model \cite{Bax}.
A slight peculiarity here is that we work on 
${\mathcal P}$, which implies an infinite system 
{}from the onset under a fixed boundary condition.
The Bethe ansatz result is 
adapted to such a circumstance as follows:
\begin{align*}
&T_m(z) \vert \xi_1, \ldots, \xi_N \rangle_B
= \lambda_m(z,\xi_1) \cdots \lambda_m(z,\xi_N) 
\vert \xi_1, \ldots, \xi_N \rangle_B,\\
& \vert \xi_1, \ldots, \xi_N \rangle_B
= \sum_{i_1 < \cdots < i_N}C_{i_1,\ldots,i_N}(\xi_1,\ldots, \xi_N)
\vert i_1, \ldots, i_N \rangle,\\
&C_{i_1,\ldots,i_N}(\xi_1,\ldots, \xi_N) = 
\sum_{P \in \mathfrak{S}_N}
\text{sign}(P) \bigl(\prod_{j < k}A_{P_j,P_k}\bigr)
\xi^{i_1}_{P_1}\cdots \xi^{i_N}_{P_N},\\
&A_{j,k} = q \eta_j - q^{-1}\eta_k,\quad 
\eta_i = \frac{1-q\xi_i}{\xi_i - q},\quad 
\lambda_m(z,\xi_i) = \frac{q^m + \eta_i z}{1+q^m\eta_i z},
\end{align*}
where $N$ is an arbitrary nonnegative integer, 
$\vert \cdots \rangle_B \in {\mathcal P}_N$ 
is the joint eigenvector of Bethe, and 
$\vert i_1, \ldots, i_N \rangle$ is the monic monomial 
describing the ball configuration at positions $i_1, \ldots, i_N$.
The sum over $P$ runs over the symmetric group $\mathfrak{S}_N$, and 
$\text{sign}(P) = \pm 1$ denotes the signature of $P$.
The above result holds for $q \in {\mathbb R}$ such that $-1 < q < 1$ 
and $z \in \C$ such that $\vert z q \vert < 1$.
The parameters $\xi_1, \ldots, \xi_N$ should be 
all distinct for the Bethe vector not to vanish.
They are to be taken {}from $\exp(\sqrt{-1}{\mathbb R})$ 
to match the condition (ii) in (\ref{eqa:pdef}), but 
otherwise arbitrary free {}from the Bethe equation.
One sees that $\lambda_m(z,\xi_i)$ tends to $\eta_i z$ 
in the limit $q^m \rightarrow 0$ in agreement with 
(\ref{eqa:homo}) with $n=2$.
The one particle eigenvalue $\lambda_m(z) = \lambda_m(z,\xi_i)$
satisfies the degenerate $T$ system 
$\lambda_m(zq)\lambda_m(zq^{-1}) = \lambda_{m+1}(z)\lambda_{m-1}(z)$.
Except the obvious $N=1$ case, it is not known to us whether 
the property $T(u) = u$ in Remark \ref{rema:eigen} can be deduced
{}from the Bethe ansatz result quoted here.
\begin{remark}\label{rema:inv}
In terms of $T_m(z)$ considered here and its transposition 
defined similarly to section \ref{subsec:norm}, 
Proposition \ref{pra:TT} is the $m \rightarrow \infty$ case 
of ${}^tT_m(z^{-1}) = T_m(z)^{-1}$ derivable {}from 
the inversion relation (\ref{eqa:inv}). 
\end{remark}

\section{\mathversion{bold} $D^{(1)}_n$ case}\label{sec:D}
\subsection{\mathversion{bold} $R$ matrix $R(z)$}
\label{subsec:R11d}

Let $J = \{1,2,\cdots, n, -n, -n+1, \cdots -1 \}$ be the set 
equipped with an order
$1 \prec 2 \prec \cdots \prec n \prec -n \prec \cdots \prec -2 \prec -1$.
In the following, elements of $2n \times 2n$ matrices
with indices {}from $J$ are arranged in the increasing order with respect to
$\prec$ from the top left.
We use the notation
\begin{equation}
\xi = q^{2n-2},\qquad 
\bar i =
  \begin{cases} 
    i  & i>0, 
    \\
    i+2n &  i<0.
  \end{cases}
\end{equation}
Let $V = \oplus_{\mu \in J} \C v_{\mu}$ be 
the vector representation of $U_q(D_n^{(1)})$.
The $R$ matrix $R(z) \in {\rm End }(V \ot V)$ was obtained in \cite{B,J}.
Here we start with the following convention:
\begin{align}
  \label{Dn-R}
  \begin{split}
  R(z) &= a(z) \sum_{k}E_{k k} \otimes E_{k k}
  +
  b(z) \sum_{j \neq k} 
  E_{j j} \otimes E_{k k} +
  c(z)\left(z\sum_{j \prec k }+\sum_{j \succ k}\right)
   E_{kj} \otimes E_{jk}  \\ 
   & ~~~~~~~~ +
  (z-1)(1-q) \sum_{j,k} f_{jk}(z)
  E_{j k} \otimes E_{-j\; -k},
  \end{split}
\end{align}
where the sums extend over $J$ and 
$E_{ij}v_k = \delta_{jk}v_i$.
\begin{align}
  \label{Dn-BW}
  \begin{split}
  &a(z) = (1-q^2z)(1-\xi z), ~
  b(z) = q(1-z)(1-\xi z), ~
  c(z) = (1-q^2)(1-\xi z),
  \\
  &f_{jk}(z) = 
  \begin{cases}
     q + \xi z & j=k,
     \\
     (1+q) (-1)^{j+k}q^{\bar{k}-\bar{j}}
     & j \prec k,
     \\
    (1+q) (-1)^{j+k}q^{\bar{k}-\bar{j}}\xi z
    & j \succ k.
  \end{cases}
  \end{split}
\end{align}
The $R$ matrix satisfies the Yang-Baxter equation.

We denote by 
$\sigma$ the automorphism of $V$ acting as 
$\sigma v_{\pm 1} = v_{\mp 1},\; 
\sigma v_{\pm n} = v_{\mp n}$, and 
$\sigma v_\mu = v_\mu$ for $\mu \neq \pm 1, \pm n$.

\subsection{\mathversion{bold} Fusion $R$ matrix and its limit}
\label{subsec:Rmd}
As the $A_{n-1}^{(1)}$ case,
we set $V_1 = V$ and realize the space $V_m$ of the $m$ fold 
$q-$symmetric tensors as the quotient
$V^{\otimes m}/A$, where 
$A = \sum_j V^{\otimes j} \ot {\rm Im}PR(q^{-2}) \ot V^{\ot m-2-j}$.
The basis of ${\rm Im}PR(q^{-2})$ can be taken as 
\begin{align*}
  &v_i\ot v_j - q v_j \ot v_i, \text{ for } i \prec j, \; i \neq \pm j,
  \\
  &v_1 \ot v_{-1} - q^2 v_{-1}\ot v_1,\quad v_n \ot v_{-n} - v_{-n}\ot v_n,
  \\
  &v_j \ot v_{-j} - v_{-j} \ot v_j 
  - qv_{-j-1}\ot v_{j+1} + q^{-1}v_{j+1}\ot v_{-j-1},
  \text{ for } 1 \le j \le n-1.
\end{align*}
A vector of the form 
$v_{i_1} \ot v_{i_2} \ot \cdots \ot v_{i_m}$ is called 
normal ordered if $-1 \succeq i_1 \succeq \cdots 
\succeq i_m \succeq 1$ and 
the sequence $i_1, \ldots, i_m$  does not 
contain the letters $n$ and $-n$ simultaneously.
The set of normal ordered vectors
$v_{i_1} \ot v_{i_2} \ot \cdots \ot v_{i_m} \mod A$ form the 
basis of $V_m$.
We label them as 
$x=[x_1,\ldots,x_n, x_{-n},\ldots,x_{-1}]$,
where $x_i \in \Z_{\ge 0}$ is the number of the letter 
$i$ in the sequence $i_1, \ldots, i_m$. 
Thus $x_1 + \cdots + x_{-1} = m$ and $x_n x_{-n} = 0$ hold
in accordance with the label in \cite{KKM}. 
In $V^{\ot m}$ normal ordering is done according to the local rule 
$\mod \hbox{Im} PR(q^{-2})$:
\begin{equation}\label{eqd:no}
\begin{split}
&v_1 \ot v_{-1} = q^2 v_{-1}\ot v_1,\quad 
v_i\ot v_j =  q v_j \ot v_i \quad i \prec j, \; i \neq \pm j, \\
&v_j \ot v_{-j} = q^2 v_{-j}\ot v_j - (1-q^2)\sum_{i=1}^{j-1}
(-q)^{j-i}v_{-i}\ot v_i\quad 2 \le j \le n-1, \\
&v_n \ot v_{-n} =  v_{-n}\ot v_n = 
-\sum_{i=1}^{n-1} (-q)^{n-i}v_{-i}\ot v_i. 
\end{split}
\end{equation}
Then the fusion $R$ matrix $R^{(m,1)}(z)$ is 
the restriction of the operator (\ref{eqd:Rcomp}) 
to ${\rm End}(V_m \ot V)$. For $x \in V_m $ and $\mu \in J$ we set 
\begin{equation}
R^{(m,1)}(z)(x \ot v_\mu) 
= \sum_{\nu \in J, y \in V_m} 
w_{\mu \nu}[x \vert y](y \ot v_\nu).
\end{equation}
Due to the weight conservation the matrix element 
$w_{\mu \nu}[x \vert y]$ is zero unless 
\begin{equation}\label{eqd:wt}
{\rm wt }(x) + {\rm wt }(v_\mu) = 
{\rm wt }(y) + {\rm wt }(v_\nu),
\end{equation}
where the weights may be regarded as elements in $\Z^n$ by
\begin{equation}\label{eqd:wtdef}
\begin{split}
&{\rm wt }([x_1,\ldots,x_n, x_{-n},\ldots,x_{-1}]) = 
(x_1-x_{-1},\ldots, x_n - x_{-n}),\\
&{\rm wt }(v_\mu) = (0\ldots,0,
\overset{\vert \mu \vert{\rm th}}{\pm 1},0,\ldots,0)\;\;
\text{ for } \pm \mu >0.
\end{split}
\end{equation}
Leaving the calculation of 
$w_{\mu \nu}[x \vert y]$ in general case aside, 
we present the result for the limit 
\begin{equation}\label{eqd:Wlim}
  W_{\mu \nu}[x \vert y] := 
  \lim_{x_{-n} \rightarrow \infty} w_{\mu \nu}[x \vert y].
\end{equation}
Note that one necessarily has $x_n = y_n = 0$ by the weight reason.
Therefore $x$ appearing in $W_{\mu \nu}[x \vert y]$ is 
to be understood as the array 
$(x_1,\ldots, x_{n-1}, x_{-n+1}, \ldots, x_{-1})$ that does not contain the 
$\pm n$ components, and the same applies to $y$ as well.
For positive integers $j$ and $k$ such that $j \le k$ 
we use the symbols
\begin{equation*}
x_{j,k} = x_j + x_{j+1} + \cdots + x_k,
\quad 
x_{-j, -k} = x_{-j} + x_{-j-1} + \cdots + x_{-k}.
\end{equation*}
They are to be understood as zero for $j>k$.
Derivation of $W_{\mu \nu}[x \vert y]$ is 
outlined in Appendix \ref{appD:W}.
We summarize the result in
\begin{proposition}
  \label{prd:W}
Suppose $j,k,l \in \{1, 2, \ldots, n-1\}$. 
The nonzero matrix elements $W_{\mu \nu}[x \vert y]$ 
are exhausted by the following list:
\begin{align*}
  \begin{split}
  &W_{\pm j,\pm j}[x \vert x] = -zq^{x_j + x_{-j}+1},
  \\
  &W_{j<k}[x \vert x-(-j)+(-k)] = 
  (-1)^{j+k} z(1-q^{2x_{-j}})q^{k-j+x_j + x_{-j-1,-k+1}},
  \\
  &W_{j>k}[x \vert x+(j)-(k)] =
  z(1-q^{2x_k})q^{x_{k+1,j-1} + x_{-k}},
  \\
  &W_{j,k}[x\vert x-(l)-(-l)+(j)+(-k)]_{l < \min(j,k)} \\ 
  & ~~ = (-1)^{k+l+1} z(1-q^{2x_l})(1-q^{2x_{-l}})
  q^{k-l-1+x_{l+1,j-1}+x_{-l-1,-k+1}},
  \end{split}
\end{align*}
\begin{align*}
  &W_{-j>-k}[x\vert x+(-j)-(-k)] =
  z(1-q^{2x_{-k}})q^{x_j + x_{-j-1,-k+1}},
  \\
  &W_{-j<-k}[x\vert x+(k)-(j)] =
  (-1)^{j+k} z(1-q^{2x_{j}})q^{j-k+x_{k+1,j-1} + x_{-k}},
  \\
  &W_{-j,-k}[x\vert x+(l)+(-l)-(j)-(-k)]_{l < \min(j,k)} \\
  & ~~ = 
  (-1)^{j+l+1} z(1-q^{2x_j})(1-q^{2x_{-k}})
  q^{j-l-1+x_{l+1,j-1} + x_{-l-1,-k+1}},
\end{align*}
\begin{align*}
  \begin{split}
  &W_{-j,k}[x\vert x-(j)+(-k)] =
  (-1)^{j+k} z^2(1-q^{2x_j})q^{j+k-2+x_{1,j-1}+x_{-1,-k+1}},
  \\
  &W_{j,-k}[x \vert x+(j)-(-k)] = 
  (1-q^{2x_{-k}})q^{x_{1,j-1}+x_{-1,-k+1}},
  \end{split}
\end{align*}
\begin{align*}
  &W_{n,k}[x\vert x-(-n)+(-k)] = 
  (-1)^{n+k} z^2 q^{n+k-2+x_{1,n-1}+x_{-1,-k+1}},
  \\
  &W_{n,-k}[x \vert x-(-n)+(k)] =
  (-1)^{n+k} zq^{n-k+x_{k+1,n-1}+x_{-k}},
  \\
  &W_{n,-k}[x \vert x+(l)+(-l)-(-n)-(-k)]_{l < k} \\
  & ~~= (-1)^{n+l+1} z(1-q^{2x_{-k}})q^{n-l-1+x_{l+1,n-1}+x_{-l-1,-k+1}},
\end{align*}
\begin{align*}
  &W_{-n,k}[x \vert x+(-n)-(k)] = 
  z(1-q^{2x_k})q^{x_{k+1,n-1}+x_{-k}},
  \\
  &W_{-n,k}[x \vert x-(l)-(-l)+(-n)+(-k)]_{l<k}\\
  & ~~= (-1)^{k+l+1} z(1-q^{2x_l})(1-q^{2x_{-l}})
  q^{k-l-1+x_{l+1,n-1}+x_{-l-1,-k+1}},
  \\
  &W_{-n,-k}[x \vert x+(-n)-(-k)] = 
  (1-q^{2x_{-k}})q^{x_{1,n-1}+x_{-1,-k+1}},
\end{align*}
\begin{align*}
  &W_{j,n}[x \vert x-(-j)+(-n)] = 
  (-1)^{j+n}z(1-q^{2x_{-j}})q^{n-j+x_j+ x_{-j-1,-n+1}},
  \\
  &W_{j,n}[x \vert x-(l)-(-l)+(j)+(-n)]_{l<j}\\
  &~~= (-1)^{l+n+1} z(1-q^{2x_l})(1-q^{2x_{-l}})
  q^{n-l-1+x_{l+1,j-1}+x_{-l-1,-n+1}},
  \\
  &W_{-j,n}[x \vert x-(j)+(-n)] =
  (-1)^{j+n} z^2(1-q^{2x_j})q^{n+j-2+x_{1,j-1}+x_{-1,-n+1}},
\end{align*}
\begin{align*}
  &W_{j,-n}[x \vert x+(j)-(-n)] =
  q^{x_{1,j-1}+x_{-1,-n+1}},
  \\
  &W_{-j,-n}[x \vert x-(-n)+(-j)] =
  zq^{x_j + x_{-j-1,-n+1}},
  \\
  &W_{-j,-n}[x \vert x-(j)-(-n)+(l)+(-l)]_{l<j}\\
  & ~~= (-1)^{j+l+1} z(1-q^{2x_j})q^{j-l-1+x_{l+1,j-1}+x_{-l-1,-n+1}},
\end{align*}
\begin{align*}
  &W_{n,n}[x \vert x] = z^2q^{2n-2+x_{1,n-1}+x_{-1,-n+1}},
  \\
  &W_{-n,-n}[x \vert x] = q^{x_{1,n-1}+x_{-1,-n+1}},
  \\
  &W_{n,-n}[x \vert x-2(-n)+(l)+(-l)] 
  = (-1)^{n+l+1} zq^{n-l-1+x_{l+1,n-1}+x_{-l-1,-n+1}},
  \\
  &W_{-n,n}[x \vert x+2(-n)-(l)-(-l)] \\
  &~~= (-1)^{n+l+1} z(1-q^{2x_l})(1-q^{2x_{-l}})
     q^{n-l-1+x_{l+1,n-1}+x_{-l-1,-n+1}}.
\end{align*}
\end{proposition}
\noindent
Here the notation 
$y = x + (l)+(-l)-(j)-(-k)$ for example means
that $y$ is obtained {}from $x$ by setting 
$x_l \rightarrow x_l + 1, x_{-l} \rightarrow x_{-l} + 1, 
x_j \rightarrow x_j - 1, x_{-k} \rightarrow x_{-k} - 1$.
Since $x_{-n}$ becomes irrelevant 
in the limit (\ref{eqd:Wlim}), $(-n)$ in the argument 
of $W_{\mu \nu}$ may just be dropped.
It has been included in the above formulas 
as a reminder of the conservation of the 
number of components.
The matrix elements of the form 
$W_{\mu \nu}[x \vert x - (\lambda) \pm \cdots]$ 
with any $\lambda \in \{\pm 1, \ldots, \pm(n-1) \}$ contain the factor 
$1-q^{2x_{\lambda}}$ as they should.

\subsection{\mathversion{bold} $L$ operator $L(z)$}
\label{subsec:Ld}

We consider the Weyl algebra generated by 
$P^{\pm 1}_\mu, Q^{\pm 1}_\mu$ with 
$\mu \in J \setminus \{\pm n\}$ under the same relation
as (\ref{eqa:pqcom}).
The subalgebra of the Weyl algebra 
generated by $P_\mu, Q_\mu$ and 
$R_\mu=Q^{-1}_{\mu}(1-a_\mu P^2_{\mu})$ 
with $\mu \in J \setminus \{\pm n\}$ will again be denoted by 
${\mathcal A}$, where $a_\mu$ is a parameter.
We define the $L$ operator 
$L(z)=(L_{\mu\nu}(z))_{\mu,\nu \in J}
\in {\mathcal A} \ot {\rm End }(V)$ so that 
$L_{\mu \nu}(z) \in {\mathcal A}$ with $\forall a_\mu = 1$ 
becomes the operator version of
$W_{\nu \mu}[x \vert y]$ in Proposition \ref{prd:W}.
See (\ref{eqd:LW}).
To present it explicitly, we assume 
$1 \le j,k,l \le n-1$
in this subsection.
We set $P'_\mu = -qa_\mu P_\mu$ and use the symbols 
\begin{align*}
&P_{j,k} = P_j P_{j+1} \cdots P_{k},\quad 
P_{-j,-k} = P_{-j}P_{-j-1} \cdots P_{-k},\\
&P'_{j,k} = P'_j P'_{j+1} \cdots P'_{k},\quad 
P'_{-j,-k} = P'_{-j}P'_{-j-1} \cdots P'_{-k}
\end{align*}
for $j \le k$. For $j > k$ they should be understood as $1$.
Then $L_{\mu\nu}(z) \in {\mathcal A}$ reads as follows:
\begin{align*}
&L_{jj}(z) = zP'_jP_{-j} + 
z\sum_{l=1}^{j-1}R_{-l}P'_{-l-1,-j+1}Q_{-j}R_lP_{l+1,j-1}Q_j,\\
&L_{-j,-j}(z) = zP_jP'_{-j} +
z\sum_{l=1}^{j-1}Q_{-l}P_{-l-1,-j+1}R_{-j}Q_lP'_{l+1,j-1}R_j,\\
&L_{k>j}(z) = zR_{-j}P'_{-j-1,-k+1}Q_{-k}P'_j +
z\sum_{l=1}^{j-1}R_{-l}P'_{-l-1,-k+1}Q_{-k}R_lP_{l+1,j-1}Q_j,\\
&L_{k<j}(z) = zP_{-k}R_kP_{k+1,j-1}Q_j + 
z\sum_{l=1}^{k-1}R_{-l}P'_{-l-1,-k+1}Q_{-k}R_lP_{l+1,j-1}Q_j,
\end{align*}
\begin{align*}
&L_{-k < -j}(z) = zQ_{-j}P_{-j-1,-k+1}R_{-k}P_j +
z\sum_{l=1}^{j-1}Q_{-l}P_{-l-1,-k+1}R_{-k}Q_lP'_{l+1,j-1}R_j,\\
&L_{-k > -j}(z) = zP'_{-k}Q_kP'_{k+1,j-1}R_j +
z\sum_{l=1}^{k-1}Q_{-l}P_{-l-1,-k+1}R_{-k}Q_lP'_{l+1,j-1}R_j,\\
&L_{k,-j}(z) = z^2P'_{-1,-k+1}Q_{-k}P'_{1,j-1}R_j,\\
&L_{-k,j}(z) = P_{-1,-k+1}R_{-k}P_{1,j-1}Q_j,
\end{align*}
\begin{align*}
&L_{k,n}(z) = z^2P'_{-1,-k+1}Q_{-k}P'_{1,n-1},\\
&L_{-k,n}(z) = zP'_{-k}Q_kP'_{k+1,n-1} +
z\sum_{l=1}^{k-1}Q_{-l}P_{-l-1,-k+1}R_{-k}Q_lP'_{l+1,n-1},\\
&L_{k,-n}(z) = zP_{-k}R_kP_{k+1,n-1} +
z\sum_{l=1}^{k-1}R_{-l}P'_{-l-1,-k+1}Q_{-k}R_lP_{l+1,n-1},\\
&L_{-k,-n}(z) = P_{-1,-k+1}R_{-k}P_{1,n-1},
\end{align*}
\begin{align*}
&L_{n,j}(z) = zR_{-j}P'_{-j-1,-n+1}P'_j + 
z\sum_{l=1}^{j-1}R_{-l}P'_{-l-1,-n+1}R_lP_{l+1,j-1}Q_j,\\
&L_{n,-j}(z) = z^2P'_{-1,-n+1}P'_{1,j-1}R_j,\\
&L_{-n,j}(z) = P_{-1,-n+1}P_{1,j-1}Q_j,\\
&L_{-n,-j}(z) = zP_jQ_{-j}P_{-j-1,-n+1} +
z\sum_{l=1}^{j-1}Q_{-l}P_{-l-1,-n+1}Q_lP'_{l+1,j-1}R_j,
\end{align*}
\begin{align*}
&L_{n,n}(z) = z^2P'_{1,n-1}P'_{-1,-n+1},\\
&L_{-n,n}(z) = z\sum_{l=1}^{n-1}Q_lP_{-l-1,-n+1}Q_{-l}P'_{l+1,n-1},\\
&L_{n,-n}(z) = z\sum_{l=1}^{n-1}R_{-l}P'_{-l-1,-n+1}R_lP_{l+1,n-1},\\
&L_{-n,-n}(z) = P_{1,n-1}P_{-1,-n+1}.
\end{align*}
In these formulas, the operators  $P_\mu, Q_\mu, R_\mu$ and $P'_\mu$ 
appearing in a single summand 
always have distinct indices hence their ordering does not matter. 

\subsection{\mathversion{bold} Factorization of $L(z)$}
\label{subsec:facLd}

For $\mu \in J \setminus \{\pm n\}$, let
$K_\mu = ((K_\mu)_{\lambda,\nu})_{\lambda,\nu \in J} 
\in {\mathcal A}\ot{\rm End }(V)$ be the operator having the 
elements
\begin{equation}\label{eqd:kdef}
  \begin{split}
  &(K_\mu)_{-n,\mu} = (K_\mu)_{-\mu, n} = Q_\mu,
  \\
  &(K_\mu)_{\mu,-n} = (K_\mu)_{n,-\mu} = R_\mu,
  \\
  &(K_\mu)_{-n,-n} = (K_\mu)_{-\mu,-\mu} = P_\mu,
  \\
  &(K_\mu)_{\mu, \mu} = (K_\mu)_{n, n} = P'_\mu, 
  \\
  &(K_\mu)_{\nu,\nu} = 1 \quad \nu \neq \pm\mu, \pm n.
  \end{split}
\end{equation}
All the other elements are zero.
Here $R_\mu = Q^{-1}_{\mu}(1-a_\mu P^2_{\mu})$ and 
$P'_\mu = -qa_\mu P_\mu$ as in section \ref{subsec:Ld}.
We also introduce $S_\mu, \bar{S}_\mu \in {\mathcal A}\ot {\rm End }(V)$ 
for $\mu = 0,\ldots, n$ as follows. 
First we specify 
$S_1, \ldots, S_{n-1}$ by 
\begin{align}
  \begin{split}
  &(S_\mu)_{\mu,\mu} = (S_\mu)_{-\mu-1,-\mu-1} = Q_\mu,
  \\
  &(S_\mu)_{\mu+1,\mu+1} = (S_\mu)_{-\mu,-\mu} = R_\mu,
  \\
  &(S_\mu)_{\mu,\mu+1} = (S_\mu)_{-\mu-1,-\mu} = P_\mu,
  \\
  &(S_\mu)_{\mu+1,\mu} = (S_\mu)_{-\mu,-\mu-1} 
  = P'_\mu, 
  \\
  &(S_\mu)_{\nu,\nu} = 1 \quad \nu \neq \pm \mu, \pm (\mu+1),
  \end{split}
\end{align}
where the other elements are zero.
Then $\bar{S}_{\mu} \in {\mathcal A}\ot {\rm End }(V)$ 
with $1 \le \mu \le n-1$ is obtained {}from $S_\mu$ by replacing 
$P_\mu, Q_\mu, R_\mu$ and $P'_\mu$ with 
$P_{-\mu}, Q_{-\mu}, R_{-\mu}=Q^{-1}_{-\mu}(1-a_{-\mu}P^2_{-\mu})$ 
and $P'_{-\mu} = -qa_{-\mu}P_{-\mu}$, respectively.
Finally the remaining ones are determined by
\begin{equation}\label{eqd:s01}
S_0 = \sigma S_1 \sigma, \quad
S_n = \sigma S_{n-1} \sigma, \quad 
\bar{S}_0 = \sigma \bar{S}_1 \sigma, \quad
\bar{S}_n = \sigma \bar{S}_{n-1} \sigma,
\end{equation}
where $\sigma = \sigma^{-1}$ 
is defined in the end of section \ref{subsec:R11d}.

The operators $K_\mu$ and $S_\nu, \bar{S}_\nu$ are 
connected via a gauge transformation 
analogous to (\ref{eqa:sdef1}).
To explain it we prepare the Weyl group 
operators $\sigma_0, \ldots, \sigma_n \in 
{\rm End }(V)$ which act as identity except
\begin{alignat*}{2}
&\sigma_0: v_1 \leftrightarrow v_{-2}, & &v_{-1} \leftrightarrow v_2,\\
&\sigma_i: v_i \leftrightarrow v_{i+1}, & &v_{-i} \leftrightarrow v_{-i-1}
\quad 1 \le i \le n-1,\\
&\sigma_n: v_{n-1} \leftrightarrow v_{-n}, & &\quad 
v_{-n+1} \leftrightarrow v_{n}.
\end{alignat*}
In terms of the sequences
\begin{align*}
&(i_{2n-2},\ldots, i_2, i_1 ) = 
(n, n-2, n-3, \ldots, 2, 0, 1, 2, \ldots, n-2, n),\\
&(\mu_{2n-2},  \ldots, \mu_2, \mu_1) =
(-n+1, \ldots, -2, -1, 1, 2, \ldots,  n-1),
\end{align*}
the gauge transformation is given by
\begin{equation}\label{eqd:ks}
K_{\mu_k} = \begin{cases}
\sigma_{i_1}\cdots \sigma_{i_k}S_{i_k}\sigma_{i_{k-1}}\cdots \sigma_{i_1}
& 1 \le k \le n-1,\\
\sigma_{i_1}\cdots \sigma_{i_k}
\bar{S}_{i_k}\sigma_{i_{k-1}}\cdots \sigma_{i_1}
& n \le k \le 2n-2.
\end{cases}
\end{equation}
We note the relations
\begin{equation}\label{eqd:sigS}
\begin{split}
&\sigma = \sigma_{i_1} \cdots \sigma_{i_{2n-2}},\\
&\sigma S_i \sigma = S_i, \quad 
\sigma \bar{S}_i \sigma = \bar{S}_i \quad 1 \le i \le n-1.
\end{split}
\end{equation}

Define the diagonal matrices 
\begin{align}
&d(z) =   z\hbox{ diag}(z^{-1}\overbrace{1, \ldots, 1}^{2n-2},z),
\nonumber\\
&D(z) = \sigma_{i_1}\cdots \sigma_{i_{n-1}}d(z)
\sigma_{i_{n-1}}\cdots \sigma_{i_1} 
= z\hbox{ diag}(\overbrace{1, \ldots, 1}^{n-1},z,z^{-1}, 
              \overbrace{1, \ldots, 1}^{n-1}). \label{eqd:dd}
\end{align}
\begin{proposition}
\label{prd:facL}
The $L$ operator in section \ref{subsec:Ld} is factorized as
\begin{equation*}
  L(z) = K_{-n+1} \cdots K_{-1} D(z) K_{1} \cdots K_{n-1}.
\end{equation*}
Equivalently it is also expressed as
\begin{align*}
L(z) &= \sigma \bar{S}_{i_{2n-2}} \cdots \bar{S}_{i_{n}} d(z) 
S_{i_{n-1}} \cdots S_{i_1}\\
&= \bar{S}_{n-1} \bar{S}_{n-2} 
         \cdots \bar{S}_{2}\bar{S}_{1}\sigma
         d(z)
         S_1 S_2 \cdots S_{n-2} S_{n}.
\end{align*}
\end{proposition}
The equivalence of the first and the second expressions is 
due to (\ref{eqd:ks}) and (\ref{eqd:dd}).
The second one and the third are connected by
(\ref{eqd:s01}) and (\ref{eqd:sigS}).
The first expression is proved in Appendix \ref{app:LK}.
\begin{proposition}\label{prd:RLL}
The $L$ operator and the $R$ matrix (\ref{Dn-R}) 
satisfy the same $RLL$ relation as
in Proposition \ref{pra:RLL}.
\end{proposition}
Proposition \ref{prd:RLL} is a corollary of 
Proposition \ref{prd:facL} and 
\begin{lemma}  \label{lemd:rss}
  \begin{align*}
  &R(z_2/z_1)(\sigma d(z_1) \ot \sigma d(z_2))
  =
  (\sigma d(z_1) \ot \sigma d(z_2)) R(z_2/z_1),\\
  &R(z)\stackrel{2}{S_\mu} \,\stackrel{1}{S_\mu}
  = \stackrel{1}{S_\mu} \, \stackrel{2}{S_\mu}R(z),    
 \quad 1 \le \mu \le n,
  \\  
&R(z)\stackrel{2}{\bar{S}_\mu} \,\stackrel{1}{\bar{S}_\mu}
  = \stackrel{1}{\bar{S}_\mu} \, \stackrel{2}{\bar{S}_\mu}R(z),    
 \quad 1 \le \mu \le n.
  \end{align*}
\end{lemma}
\begin{proof}
The first relation is straightforward to check.
Next consider the second relation with $1 \le \mu \le n-1$.
Comparing the $R$ matrices (\ref{eqa:r}) and (\ref{Dn-R}),
we find that the contributions proportional to 
$a(z), b(z)$ and $c(z)$ on the both sides are equal due to 
Lemma \ref{lema:rss} for $A^{(1)}_{n-1}$ case.
Thus we are to show the equality with $R(z)$ replaced with 
$\sum_{j,k} f_{jk}(z)E_{j k} \otimes E_{-j\; -k}$.
It is easily checked at $z=0$ and $z=\xi^{-1}$ for example, which 
suffices since $f_{jk}(z)$ is linear in $z$.
Then the second relation with $\mu = n$ follows
{}from $\mu=n-1$ case by using 
$S_n = (\sigma d(z))^{-1}S_{n-1} \sigma d(z)$.
The third relation can be shown similarly.
\end{proof}
As Remark \ref{rema:K2}, if $a_\mu = 1$, 
the property $K_\mu^2=\openone_{2n}$ holds for any 
$\mu \in J \setminus \{ \pm n \}$.

\subsection{\mathversion{bold}
Quantized $D^{(1)}_n$ automaton}
\label{subsec:dbbs}

Here we set up the quantized $D^{(1)}_n$ automaton.
It is a system of particles and 
antiparticles on one dimensional lattice whose 
dynamics is governed by the $L$ operator constructed 
in section \ref{subsec:Ld}.
In the limit $q \rightarrow 0$, the 
dynamics become deterministic and the system 
reduces to the $D^{(1)}_n$ automaton \cite{HKT3,HKT1}.
Since our results are parallel with those in 
subsections \ref{subsec:qbbs1} -- \ref{subsec:norm},
we shall only give a brief sketch and omit the details.

The space of states ${\mathcal P}$ is given by 
(\ref{eqa:pdef}), where $V$ is now 
understood as the $2n$ dimensional vector representation 
$V = \C v_1 \oplus \cdots \oplus \C v_{-1}$.
The condition (ii) remains the same while 
the condition (i) is replaced by 
$\sum_{k \in \Z} \vert j_k + n \vert < \infty$.

Monomials 
$\cdots \ot v_{j_{-1}}\ot v_{j_0} \ot v_{j_1} \ot \cdots$ 
can be classified according to 
the numbers $w_1,\ldots, w_{n}, w_{-n+1},\ldots, w_{-1}$ 
of occurrence of the 
letters $1, \ldots, n, -n+1, \ldots, -1$ in the set $\{j_k\}$.
Consequently one has the direct sum decomposition
${\mathcal P} = 
\oplus {\mathcal P}_{w_1,\ldots,  w_{-1}}$
analogous to (\ref{eqa:dsd}), where 
${\mathcal P}_{0,\ldots,0}= \C p_{\rm vac}$ with 
$p_{\rm vac} = \cdots \ot v_{-n} \ot v_{-n} \ot \cdots$.

The local states $v_{j_k} \in V$ is regarded as 
the $k$th box containing a particle of color $j_k$ if 
$j_k \in \{\pm 1, \ldots, \pm (n-1)\}$. Particles having 
colors with opposite signs are regarded as antiparticles of the other.
The case $j_k = -n$ is interpreted as an empty box, while 
$j_k = n$ represents a bound state of a particle and an antiparticle.

To formulate the time evolution, we assume 
$\forall a_\mu = 1$ {}from now on, and consider the space of the 
quantum carrier, namely, the 
${\mathcal  A}$ module ${\mathcal M}$ defined similarly to
(\ref{eqa:M}).
The difference now is that we need 
$2n-2$ coordinates and to set 
${\mathcal M} = \oplus \C 
[m_1, \ldots, m_{n-1}, m_{-n+1}, \ldots, m_{-1}]$.
Then the actions of $P_\mu, Q_\mu, R_\mu$ and 
$P'_\mu = -qP_\mu$ are again 
given by (\ref{eqa:M}) by simply extending the index $i$ to 
$\mu = \pm 1, \ldots, \pm(n-1)$.
By construction we have
\begin{equation}\label{eqd:LW}
L(z)(x \ot v_\mu) = \sum_{\nu \in J,\, y\in {\mathcal M}} 
W_{\mu \nu}[x \vert y](y \ot v_\nu)
\end{equation}
for $x \in {\mathcal M}$.
Here the sum over $y$ is taken 
under the constraint (\ref{eqd:wt}), where
the weight $\text{wt}$ should now be understood as 
(\ref{eqd:wtdef}) without the $n$th component.

The time evolution $T(z): {\mathcal P} \rightarrow {\mathcal P}$
is also given by the same formula (\ref{eqa:defT}), where 
$(\cdots )_{0,0}$ now signifies the element in ${\rm End }({\mathcal P})$
corresponding to the transition {}from 
$[\overbrace{0,\ldots,0}^{2n-2}]$ to itself in the ${\mathcal M}$ part.
{}From (\ref{eqd:dd}) one has 
$T(z)p = z^{w_1+\cdots+w_{n-1}+2w_n+ w_{-n+1}+\cdots+w_{-1}}T(1)p$
for $p \in  {\mathcal P}_{w_1,\ldots,  w_{-1}}$.
The power of $z$ is the total number of particles and 
antiparticles, for $v_n$ represents a bound state of a particle and 
an antiparticle. 
As it turns out, the total number is conserved, which implies 
the commutativity $T(z)T(z') = T(z')T(z)$.
We concentrate on $T=T(1)$ henceforth.

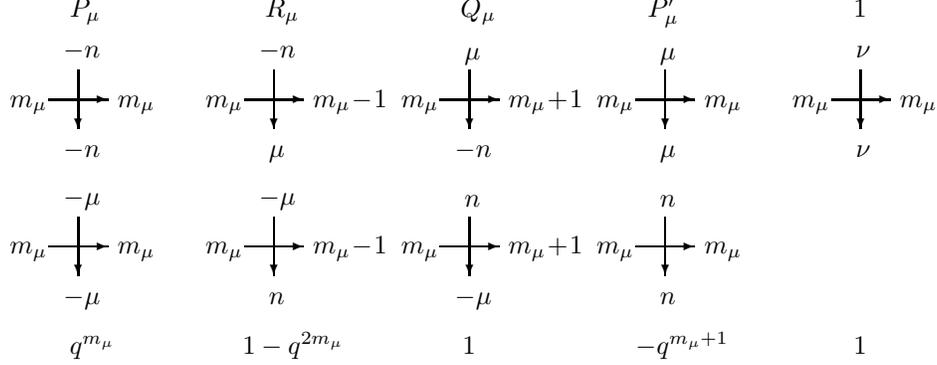
\begin{figure}[h]
\setlength{\unitlength}{1.3mm}
\caption{Diagram for $K_\mu\; (\nu \neq \pm\mu, \pm n)$}\label{fig:Kmu}
\begin{picture}(125,38)(0,-3.5)
\put(7.1,31.5){$P_\mu$}\put(27.1,31.5){$R_\mu$}\put(47.1,31.5){$Q_\mu$}
\put(66.2,31.5){$P'_\mu$}\put(87.3,31.5){$1$}
\multiput(0,15)(20,0){5}{
\put(5,8){\vector(1,0){6}}\put(8,11){\vector(0,-1){6}}
\put(1.0,7.3){${m_\mu}$}}
\multiput(0,0)(20,0){4}{
\put(5,8){\vector(1,0){6}}\put(8,11){\vector(0,-1){6}}
\put(1.0,7.3){${m_\mu}$}}

\multiput(0,0)(0,15){2}{
\put(12,7.3){${m_\mu}$}
\put(32,7.3){${m_\mu}\!-\!1$}
\put(52,7.3){${m_\mu}\!+\!1$}
\put(72,7.3){${m_\mu}$}}
\put(92,22.3){${m_\mu}$}

\put(6.5,27.3){$-n$}\put(6.5,17.1){$-n$}
\put(26.5,27.3){$-n$}\put(27.5,17.1){$\mu$}
\put(47.5,27.1){$\mu$}\put(46.5,17.1){$-n$}
\put(67.5,27.1){$\mu$}\put(67.5,17.1){$\mu$}

\put(6.5,12.3){$-\mu$}\put(6.5,2.1){$-\mu$}
\put(26.5,12.3){$-\mu$}\put(27.5,2.1){$n$}
\put(47.5,12.1){$n$}\put(46.5,2.1){$-\mu$}
\put(67.5,12.1){$n$}\put(67.5,2.1){$n$}
\put(87.5,27.3){$\nu$}\put(87.5,17.1){$\nu$}

\put(7.1,-3){$q^{m_\mu}$}
\put(24.8,-3){$1-q^{2{m_\mu}}$}
\put(47.3,-3){$1$}
\put(65,-3){$-q^{{m_\mu}+1}$}
\put(87.3,-3){$1$}

\end{picture}
\end{figure}

The propagation operators ${\mathcal K}_\mu$
for $\mu = \pm 1, \ldots, \pm(n-1)$ are defined 
in the same way as \eqref{eqa:pKdef1} as the product of
$K_\mu$ acting locally.  
This time the local interaction and 
their amplitudes implied by 
(\ref{eqd:kdef}) are depicted in Fig. \ref{fig:Kmu}.

Here ${m_\mu} \in \Z_{\ge 0}$ is a coordinate in 
$[m_1,\ldots, m_{n-1},m_{-n+1},\ldots, m_{-1}] \in {\mathcal M}$, 
meaning the number of color $\mu$ particles on the carrier.
The top five diagrams are essentially the same as 
Fig. \ref{fig:Ki} for $A^{(1)}_{n-1}$ case, where 
color $\mu$ particles on the carrier (horizontal line) 
behave according to the 
presence or absence of another color $\mu$ particle in a local box.
(The empty box $-n$ here corresponds to $n$ in $A^{(1)}_{n-1}$ case.)
The bottom four vertices are new.
The second one there is the pair annihilation of a color 
$\mu$ particle on the carrier and the antiparticle $-\mu$ 
in the box to form the bound state $n$.
The third one is the pair creation of 
$\mu$ and $-\mu$ {}from the bound state $n$.
At $q=0$ the amplitudes for $P_\mu$ with $m > 0$, 
$R_\mu$ with $m=0$  
and $P' _\mu$ vanish and the other ones
become 1. As the result they reduce to the deterministic rule 
that agrees with the one in \cite{HKT3}.

As a parallel result with Theorem \ref{th:facT}, 
the time evolution of the quantized $D^{(1)}_n$ automaton 
admits the factorization into the propagation operators.
\begin{theorem}\label{thd:facT}
\begin{equation*}
T = {\mathcal K}_{-n+1}\cdots{\mathcal K}_{-1}
{\mathcal K}_{1}\cdots{\mathcal K}_{n-1}.
\end{equation*}
\end{theorem}
This is a consequence of Proposition \ref{prd:facL}.
It extends a part of the  earlier result  
at $q=0$ based on the crystal basis theory \cite{HKT2,HKT3}, 
where the time evolutions of a class of soliton cellular automata 
were factorized.

Finally we state properties of the amplitude for $T$.
Define the transposition ${}^tT$ of $T$, the subspace 
${\mathcal P}_{\text{fin}}$ and the linear function
${\mathcal N}:  {\mathcal P}_{\text{fin}} \rightarrow \C$ 
in the same manner as section \ref{subsec:norm}.
\begin{proposition}\label{prod:matome}
Proposition \ref{pra:TT} and Proposition \ref{pra:norm1}
are both valid also for the quantized $D^{(1)}_n$ automaton.
\end{proposition}
\begin{proof}
In view of the factorization of $T$,
it is enough to show the claim for any one of 
the propagation operators, say ${\mathcal K}_1$.
Namely ${}^t{\mathcal K}_1 = {\mathcal K}^{-1}_1$ 
and ${\mathcal N}({\mathcal K}_1(p)) = {\mathcal N}(p)$.
Then without a loss of generality 
one may restrict the space of states to
${\mathcal P}_{w_1,\ldots, w_{-1}}$ with all 
$w_\mu$ being zero except $w_{\pm 1}$ and $w_n$.
Let $\pi$ be the map that embeds the local states into that for 
$A^{(1)}_1$ as $\pi(v_1) = \pi(v_n) = \bullet$ (a ball) and 
$\pi(v_{-1}) = \pi(v_{-n}) = \circ$ (an empty box), where 
we have used the notation in section \ref{subsec:norm}
for $A^{(1)}_1$.
Let further $\phi$ be the map sending the pair of local states for 
$A^{(1)}_1$ and $D^{(1)}_n$ to that for the latter as 
\begin{alignat*}{4}
\phi(\circ,v_1) &= v_{-n}, & 
\;\;\phi(\circ,v_{-1}) &= v_{-1}, & 
\;\;\phi(\circ,v_{-n}) &= v_{-n}, & 
\;\;\phi(\circ,v_n) &= v_{-1}, \\
\phi(\bullet,v_{1}) &= v_{1}, &
\phi(\bullet,v_{-1}) &= v_{n}, &
\phi(\bullet,v_{-n}) &= v_{1}, &
\phi(\bullet,v_{n}) &= v_{n}.
\end{alignat*}
The componentwise action of these maps will also be denoted by the 
same symbol. For example,
if $p = \cdots \ot v_{-n} \ot v_{-1} \ot v_{-n} \ot \cdots$ and 
$p' = \cdots \ot \circ \ot \bullet \ot \circ \ot \cdots$ in the 
corresponding position, one has 
$\pi(p) = \cdots \ot \circ \ot \circ \ot \circ \ot \cdots$ and 
$\phi(p',p) = \cdots \phi(\circ, v_{-n}) \ot
\phi(\bullet, v_{-1}) \ot
\phi(\circ, v_{-n}) \ot \cdots
= \cdots \ot v_{-n} \ot v_n \ot v_{-n} \ot \cdots$.
Denoting the propagation operator for $A^{(1)}_1$ by ${\mathcal K}^A_1$,
one has the embedding 
${\mathcal K}_1(p) = \phi\bigl({\mathcal K}^A_1(\pi(p)),p\bigr)$.
With the aid of this relation, 
the statements are reduced to the $A^{(1)}_1$ case 
established in section \ref{subsec:norm}.
\end{proof}

\appendix
\section{Proof of Proposition \ref{prd:W}}\label{appD:W}

The simplifying feature of the limit 
$x_{-n} \rightarrow \infty$ (\ref{eqd:Wlim}) is that one can decompose 
$w_{\mu\nu}[x \vert y]$ into three parts effectively.
To see this suppose 
$x \in V_m$ is in normal order
\begin{equation*}
(v_{-1})^{\ot x_{-1}} \ot \cdots \ot 
(v_{-n+1})^{\ot x_{-n+1}} 
\ot (v_{-n})^{\ot x_{-n}}
\ot (v_{n-1})^{\ot x_{n-1}} 
\ot \cdots \ot (v_{1})^{\ot x_{1}}.
\end{equation*}
Application of (\ref{eqd:Rcomp}) for $D^{(1)}_n$ to this generates 
a variety of vectors 
$y = v_{j_1} \ot \cdots \ot v_{j_m}$.
However in the limit $x_{-n} \rightarrow \infty$ under consideration,
the vectors $v_1, \ldots, v_{n}$ are not allowed to 
appear in the left side of the 
segment $v_{-n} \ot \cdots \ot v_{-n}$ 
since they acquire the factor of order $q^{x_{-n}}$ 
in the course of normal ordering. See (\ref{eqd:no}).
Similarly, $v_{-1}, \ldots, v_{-n+1}$ are forbidden to show up 
in the right side of $v_{-n} \ot \cdots \ot v_{-n}$.
In this way $W_{\mu\nu}[x \vert y]$ is effectively 
decomposed into the right, left and 
the infinitely large central parts, where 
the allowed indices are limited to 
$\{1,\ldots, n-1\}$, $\{-1,\ldots, -n+1\}$ and $-n$, respectively.

Taking the situation into account, 
we derive $W_{\mu\nu}[x \vert y]$ (\ref{eqd:Wlim}) in three steps.
In {\em Step 1}, we compute all the matrix elements 
$w_{\mu\nu}[x \vert y]$ for $x$ of the form 
$x=i^m = \overbrace{v_i \ot \cdots \ot v_i}^m$, which 
serves as a building block for general $x$.
In {\em Step 2}, 
we obtain the limits of $w_{\mu\nu}[x \vert y]$ that are
relevant to the three parts separately.
In {\em Step 3}, we glue the three parts together.

\noindent{\em Step 1}.
\begin{lemma}\label{lemd:wm}
All the matrix elements of the form 
$w_{j,k}[i^m \vert y]$ are zero except the following:
\begin{align}
\label{w-1}
&w_{i,i}[i^m \vert i^m] = (1-q^{m+1}z)(1-q^{m-1}\xi z) \quad \forall i,
\\
\label{w-2}
&w_{-i,-i}[i^m \vert i^m] = 
(q^{m-1}-z)(q^{m+1}-\xi z) \qquad \forall i,  \\
  \label{w-3}
  &w_{j,j}[i^m \vert i^m] = q(q^{m-1}-z)(1-q^{m-1}\xi z)
  \quad j \neq \pm i\, \; \forall i,
  \\
  \label{w-4}
  &w_{j,i}[i^m \vert i^{m-1},j] = (1-q^{2m})(1-q^{m-1}\xi z) \times 
  \begin{cases}
    1 & i \succ j, \; j \neq \pm i\\
    z & i \prec j, \; j \neq \pm i
  \end{cases}
  \quad \forall i,
  \\
  \label{w-5}
  &w_{-i,j}[i^m \vert -j,i^{m-1}] = 
  (-1)^{i+j+1}(1-q^{2m})(q^{m-1}-z)q^{\bar{j} + \bar{i}-2}\times
  \begin{cases} z & 1 \preceq j \prec -i \\
                \xi^{-1} & -i \prec j \preceq -1
  \end{cases}\quad \forall i,
  \\
  \label{w-6}
  &w_{-i, i}[i^m \vert i^{m-2},j, -j] =  
  (-1)^{i+j+1}
  q^{n-j-1}(1-q^{2m})(1-q^{2m-2}\xi)z \quad i = \pm n,\; 1 \le j \le n-1,
  \\
  \label{w-7}
  \begin{split}
  &w_{-i,i}[i^m \vert -i,i^{m-1}] \\
  &= \begin{cases}
     q^{m-1}(1-q^{2m})(1-q^{m-1}\xi z+q^{2i-1-m}(z-q^{m-1}))z & 
     1 \leq i \leq n-1\\
     (1-q^{2m})(1-q^{m-1}\xi z+q^{2i+1+m}\xi(z-q^{m-1})) & 
     -n+1 \preceq i \preceq -1
     \end{cases}\quad i \neq \pm n,
  \end{split}
  \\
  \label{w-8}
  \begin{split}
  &w_{-i,i}[i^m \vert i^{m-2},j,-j] \\
  &= 
  \begin{cases}
    (-1)^{i+j+1}q^{i-j-1}(1-q^{2m})(1-q^{2m-2})(1-q^{m-1}\xi z)z &
    1 \leq i \leq n-1,\;  1 \le j < i\\
    (-1)^{i+j}q^{i-j+1} \xi (1-q^{2m})(1-q^{2m-2})(q^{m-1}-z) &
    -n+1 \preceq i \preceq -1, \; 1 \le j < \vert i \vert.
  \end{cases}
  \end{split}
\end{align}
\end{lemma}
In these formulas for $w_{j,k}[i^m \vert y]$, 
$y$ should be understood as a normal ordered vector 
in $V_m$ having the specified contents of the letters.

\noindent
{\em  Sketch of the proof}.
The first four, \eqref{w-1}-\eqref{w-4}, are
straightforward to check.
The other formulas \eqref{w-5}--\eqref{w-8} 
are shown in this order by induction on $m$. 
Here we illustrate it for \eqref{w-8}.
Let us write the $R$-matrix \eqref{Dn-R} as
$
  R(z) = \sum_{i,j,k,l} 
           r[i,k;j,l](z)  E_{ji} \ot E_{lk}.
$
For simplicity $w_{j,k}[i^m \vert y](z)$ will be denoted by
$w_{j,k}[y](z)$.
We treat the case $1 \leq j < i \leq n-1$.
The result \eqref{w-8} for $m=2$ can be checked directly.
Assume \eqref{w-1}--\eqref{w-8} up to $m$.
The fusion construction leads to 
the following recursion relation for $m \geq 3$:
\begin{align*}
  w_{-i,i}&[i^{m-1},j,-j](z) a(zq^{m-2})
  \\
  =
  & ~ \underline{q} ~r[i,i;i,i](z q^m) ~w_{-i,i}[i^{m-2},j,-j](z q^{-1})
  \\
  &+ \sum_{\alpha \neq i, \alpha=j+1}^{n-1}      
     \underline{(-1)^{1+\alpha+j}(1-q^2) q^{\alpha-j+m-1}} ~ 
      r[i,\alpha;\alpha,i](z q^m) ~
     w_{-i,\alpha}[i^{m-1},-\alpha](z q^{-1})  
  \\
  &+ \underline{(-1)^{i+j+1}(1-q^2) q^{i-j+m-1}}r[i,i;i,i](z q^m) ~ 
     w_{-i,i}[i^{m-1},-i](z q^{-1})
  \\
  & + \underline{q^{m+1}}~ r[i,j;j,i](z q^m) ~
  w_{-i,j}[i^{m-1},-j](z q^{-1})
  \\
  & + r[i,-j;-j,i](z q^m) ~ w_{-i,-j}[i^{m-1},j](z q^{-1})
  \\
  & + \underline{(-1)^{n+j+1}q^{n-j+m-1}} 
    \bigl( ~r[i,-n;-n,i](z q^m) ~ w_{-i,-n}[i^{m-1},n](z q^{-1})
           \\ & ~~~~~~~~~~~~~~~~~~~~~~~ 
           + r[i,n;n,i](z q^m) ~ w_{-i,n}[i^{m-1},-n](z q^{-1})
    ~ \bigr), 
\end{align*}
where the underlined factors come from the normal ordering.
To check that \eqref{w-1}--\eqref{w-8} satisfy 
this is easy. 
\vspace{-0.57cm}\begin{flushright}$\square$\end{flushright}

\noindent
{\em Step 2}. 
As explained in the beginning of the appendix, 
we investigate the three parts that constitute the limit 
$W_{\mu\nu}[x \vert y]$ separately.
First we consider the right part.
\begin{lemma}\label{lemd:ue}
Set $w^\prime_{\mu\nu}[x \vert y] = w^\prime_{\mu\nu}[x \vert y](z)
= w_{\mu\nu}[x \vert y]/a(z)$ and $m_1 = x_{1,n-1}$.
Suppose $x$ and $y$ have the form
$x=[x_1,\ldots,x_{n-1},0,\ldots,0]$ and 
$y=[y_1,\ldots,y_{n-1},y_{-n},0,\ldots,0]$, respectively.
Then the nonzero case of the 
limit $\lim_{z \rightarrow \infty}w^\prime_{\mu\nu}[x \vert y]$ 
is given by
\begin{equation*}
  \begin{split}
  &w^\prime_{\pm j,\pm j}[x \vert x] \to q^{-m_1 \pm x_j}
  ~(1 \leq j \leq n),
  \\ 
  &w^\prime_{j,k}[x \vert x+(j)-(k)] \to -(1-q^{2 x_k})q^{-m_1-1+x_{k+1,j-1}}
  ~(1 \leq k < j \leq n-1),
  \\
  &w^\prime_{-j,-k}[x \vert x-(j)+(k)] \to 
   (-1)^{j+k}(1-q^{2 x_j})q^{-m_1+j-k-x_{j,k}}
  ~(1 \leq j < k \leq n-1),
  \\
  &w^\prime_{-n,k}[x \vert x+(-n)-(k)] \to -(1-q^{2 x_k})q^{-1-x_{1,k}}
  ~(1 \leq k \leq n-1),
  \\
  &w^\prime_{-j,n}[x \vert x-(j)+(-n)] \to (-1)^{j+n}(1-q^{2 x_j})q^{-m_1+j-n-x_{j,n-1}}
  ~(1 \leq j \leq n-1).
  \end{split}
\end{equation*}
\end{lemma}
\noindent{\em Sketch of the proof}.
We illustrate the derivation of the second case.
{}From the fusion construction one gets 
\begin{align*}
  \begin{split}
  &w^\prime_{j,k}[x \vert x+(j)-(k)](z q^{m_1-1})
  =
  q^{x_{k+1,j-1}}
  \frac{w_{j,k}[k^{x_k} \vert k^{x_k-1},j](zq^{x_k-1})}
       {a(z q^{2(x_k-1)})}
  \\
  & ~~~~~ \times
  \Bigl(
  \prod_{i=1}^{k-1} 
  \frac{w_{k,k}[i^{x_i}\vert i^{x_i}](z q^{2 x_k+2x_{1,i-1}+x_i-1})}
       {a(z q^{2x_k+2(x_{1,i}-1)})}
  \prod_{i=k+1}^{n-1} 
  \frac{w_{k,k}[i^{x_i}\vert i^{x_i}](z q^{2x_{1,i-1}+x_i-1})}
       {a(z q^{2(x_{1,i}-1)})}
  \Bigr), 
  \end{split}   
\end{align*}
where the factor $q^{x_{k+1,j-1}}$ is due to normal ordering.
Substituting \eqref{w-3} and \eqref{w-4}, one finds 
that this tends to the desired form in the limit $z \rightarrow \infty$.
\vspace{-0.57cm}\begin{flushright}$\square$\end{flushright}

Next we deal with the central part.
\begin{lemma}\label{lemd:naka}
Nonzero limit $q^{x_{-n}} \rightarrow 0$ 
of $w_{\mu \nu}[(-n)^{x_{-n}} \vert y]$ is given by
\begin{equation*}
\begin{split}
  &w_{n,n}[(-n)^{x_{-n}} \vert (-n)^{x_{-n}}] \to \xi z^2, 
  \\
  &w_{-n,-n}[(-n)^{x_{-n}} \vert (-n)^{x_{-n}}] \to 1,
  \\
  &w_{n,-n}[(-n)^{x_{-n}} \vert -j,(-n)^{x_{-n}-2},j] \to 
  (-1)^{j+n+1} q^{n-j-1}z,
  \\
  &w_{n,j}[(-n)^{x_{-n}} \vert -j,(-n)^{x_{-n}-1}] \to 
  (-1)^{j+n} q^{n+j-2}z^2, 
  \\ 
  &w_{n,-j}[(-n)^{x_{-n}} \vert (-n)^{x_{-n}-1},j] \to 
  (-1)^{j+n} q^{n-j}z,
  \\
  &w_{\pm j,\pm j}[(-n)^{x_{-n}} \vert (-n)^{x_{-n}}] \to -q z,
  \\
  &w_{j,-n}[(-n)^{x_{-n}} \vert (-n)^{x_{-n}-1},j] \to 1, 
  \\
  &w_{-j,-n}[(-n)^{x_{-n}} \vert -j,(-n)^{x_{-n}-1}] \to z,
\end{split}
\end{equation*} 
where $1 \leq j \leq n-1$.
\end{lemma}
\begin{proof}
Straightforward calculation based on Lemma \ref{lemd:wm}.
\end{proof}
Finally for the left part, the following is verified 
similarly to Lemma \ref{lemd:ue}.
\begin{lemma}\label{lemd:shita}
Suppose $x$ and $y$ have the form 
$x=[0,\ldots,0,x_{-n+1},\ldots,x_{-1}]$ and 
$y=[0,\ldots,0,y_{-n},y_{-n+1},\ldots,y_{-1}]$.
Then the nonzero case of the 
limit $\lim_{z \rightarrow 0}w_{\mu\nu}[x \vert y]$ 
is given by
\begin{align*}
  \begin{split}
  &w_{\pm j,\pm j}[x \vert x] \to q^{m_2 \pm x_{-j}}
  ~(1 \leq j \leq n),
  \\
  &w_{j,k}[x \vert x-(-j)+(-k)] \to 
   (-1)^{j+k+1}(1-q^{2 x_{-j}})q^{m_2 +k-j-1+x_{-j-1,-k+1}}
  ~(1 \leq j < k \leq n),
  \\
  &w_{-j,-k}[x \vert x+(-j)-(-k)] \to 
   (1-q^{2 x_{-k}})q^{m_2 -x_{-k,-j}}
  ~(1 \leq k < j \leq n),
  \end{split}
\end{align*}
where $m_2 = x_{-1,-n+1}$.
\end{lemma}
 
\noindent{\em Step 3}.
We demonstrate the gluing procedure with two examples.
First we derive the 4th case in Proposition \ref{prd:W},
$W_{i,l}[x \vert x+(i)-(j)-(-j)+(-l)]$.
This is calculated as the simple product of the three parts:
\begin{align*}
  \begin{split}
  &w^\prime_{i,j}[x \vert x+(i)-(j)](z q^{-m+m_1})
  w_{j,j}[(-n)^{x_{-n}} \vert (-n)^{x_{-n}}](z q^{m_1-m_2})
  \\
  & ~~~~~
  \times w_{j,l}[x^\prime \vert x^\prime-(-j)+(-l)](z q^{m-m_2}),
  \end{split}
\end{align*}
which is nonzero for $1 \leq j \leq \min(i,l)$.
For $j<i<l$, it is calculated by multiplying 
the second one in Lemma \ref{lemd:ue},
the 6th of Lemma \ref{lemd:naka} and the second of Lemma \ref{lemd:shita},
leading to 
\begin{align*}
  &-(1-q^{2 x_j})q^{-m_1-1+x_{j+1,i-1}}\times
  (-z q^{1+m_1-m_2}) \times
  (-1)^{j+l+1}(1-q^{2 x_{-j}})q^{m_2+l-j-1+x_{-j-1,-l+1}}
  \\
  &~~ = 
  (-1)^{j+l+1} z (1-q^{2 x_j}) (1-q^{2 x_{-j}})
  q^{l-j-1+x_{j+1,i-1}+x_{-j-1,-l+1}}.         
\end{align*}
This agrees with the sought result.
Second we consider the 9th case in Proposition \ref{prd:W},
$W_{i,-k}[x \vert x+(i)-(-k)]$.
This matrix element is obtained by 
collecting several contributions as
\begin{align*}
  \begin{split}
  &\Bigl(
  \underline{q^{x_{i+1,n-1}}}
  w^\prime_{i,i}[x \vert x](z q^{-m+m_1}) \,
  w_{i,-n}[(-n)^{x_{-n}} \vert (-n)^{x_{-n}-1},i](z q^{m_1-m_2})
  \\
  & ~~~~~
  +
  \sum_{j=1}^{i-1}
  \underline{q^{x_{j+1,n-1}+1}} 
  w^\prime_{i,j}[x \vert x+(i)-(j)](z q^{-m+m_1}) \,
  w_{j,-n}[(-n)^{x_{-n}} \vert (-n)^{x_{-n}-1},j](z q^{m_1-m_2}) 
  \Bigr)
  \\ 
  & ~~~~~
  \times
  w_{-n,-k}[x^\prime \vert x^\prime-(-k)+(-n)](z q^{m-m_2}),
  \end{split}
\end{align*}
where we have set $x=[x_1,\ldots, x_{n-1},0,\ldots,0]$ and 
$x'=[0,\ldots,0,x_{-n+1},\ldots,x_{-1}]$. 
The underlined factors come from normal ordering.
In the limit $x_{-n} \rightarrow \infty$, this is evaluated 
by using the first two of Lemma \ref{lemd:ue},
the 7th of Lemma \ref{lemd:naka} and the last of 
Lemma \ref{lemd:shita} as
\begin{align*}
  &\Bigl( q^{x_{i,n-1}} 
         - \sum_{j=1}^{i-1}(1-q^{2x_j})q^{x_{j+1,i-1}+x_{j+1,n-1}}
  \Bigr)
  (1-q^{2x_{-k}}) q^{m_2-x_{-k,-n+1} -m_1}.
\end{align*}
The sum  leads to the result 
$(1-q^{2x_{-k}}) q^{x_{1,i-1}+x_{-1,-k+1}}$. 
\vspace{-0.57cm}\begin{flushright}$\square$\end{flushright}


\section{Proof of Proposition \ref{prd:facL}}\label{app:LK}

Let $L_n\Bigl[\begin{matrix} P^\prime & R \\ Q & P \end{matrix}\Bigr]$
be the $L$ operator $L(z)$ for $A_{n-1}^{(1)}$ with $z=1$
defined in (\ref{eqa:L-elements}). 
The $L$ operator with $P_i$ and $P'_{i}$ interchanged for all 
$i \in \{1,\ldots, n-1\}$ will be denoted by 
$L_n\Bigl[\begin{matrix} P & R \\ Q & P^\prime \end{matrix}\Bigr]$.
A similar convention is applied also for the other interchanges like 
$R_i \leftrightarrow Q_i$, etc.
A matrix $\bar{L}_n[\cdots]$ is the one 
obtained {}from $L_n[\cdots]$ 
by changing $X_i \,(X=P, P', Q, R)$  into $X_{-i}$ 
for all $i  \in \{1,\ldots, n-1\}$.
Matrices $L_n^+[\cdots]$ and $\bar{L}^+_n[\cdots]$ 
are the ones obtained {}from $L_n[\cdots]$ and 
$\bar{L}_n[\cdots]$ respectively by the replacement 
$X_{\pm i} \rightarrow X_{\pm(i+1)}$ for 
all $i  \in \{1,\ldots, n-1\}$.
For any square matrix $M$ we let 
$\Tilde{M}$ denote the one obtained by 
reversing the order of rows and columns simultaneously.
\begin{lemma}\label{lemd:ind}
\begin{align*}
 &\begin{pmatrix}
  P_1^\prime & & R_1\\
  & \openone_{n-1}\\
  Q_1 & & P_1
  \end{pmatrix}
  \begin{pmatrix}
  1 \\
  & L_n^+ \Bigl[\begin{matrix} P^\prime & R \\ Q & P
                  \end{matrix}\Bigr]\\ 
  \end{pmatrix}
  =
L_{n+1}\Bigl[\begin{matrix} P^\prime & R \\ Q & P \end{matrix}\Bigr],\\
   &\begin{pmatrix}
  P_1^\prime & & R_1\\
  & \openone_{n-1}\\
  Q_1 & & P_1
  \end{pmatrix}
  \begin{pmatrix}
  \Tilde{L}_n^+ \Bigl[\begin{matrix} P & Q \\ R & P^\prime
                  \end{matrix}\Bigr]\\ 
  & 1 
  \end{pmatrix}
  =
  \Tilde{L}_{n+1}\Bigl[\begin{matrix} P & Q \\ R & P^\prime 
                         \end{matrix}\Bigr],
  \\
   &\begin{pmatrix}
  1 \\
  & ^t\bar{L}_n^+ \Bigl[\begin{matrix} P & R \\ Q & P^\prime
                  \end{matrix}\Bigr]\\ 
  \end{pmatrix}
  \begin{pmatrix}
  P_{-1} & & Q_{-1}\\
  & \openone_{n-1}\\
  R_{-1} & & P_{-1}^\prime
  \end{pmatrix}
  =
  ~ ^t\bar{L}_{n+1}\Bigl[\begin{matrix} P & R \\ Q & P^\prime 
                         \end{matrix}\Bigr]. 
  \\
   &\begin{pmatrix}
  ^t\Tilde{\bar{L}}_n^+ \Bigl[\begin{matrix} P^\prime & Q \\ R & P
                              \end{matrix}\Bigr]\\ 
  & 1 \\
  \end{pmatrix}
  \begin{pmatrix}
  P_{-1} & & Q_{-1}\\
  & \openone_{n-1}\\
  R_{-1} & & P_{-1}^\prime
  \end{pmatrix}
  =
  ~ ^t\Tilde{\bar{L}}_{n+1}\Bigl[\begin{matrix} P^\prime & Q \\ R & P
                                 \end{matrix}\Bigr].
  \end{align*}
Here ${}^t$ means the transposition.
\end{lemma}
\begin{proof}
The first relation is just (\ref{eqa:kpro}).
The second relation is obtained {}from the first one 
by taking $\,\tilde{\;}\,$ and the interchanges 
$P \leftrightarrow P', Q \leftrightarrow R$.
See Remark \ref{rema:com}.
The third relation follows {}from the first one 
by ${}^t\,\bar{\;}\,$ and $P \leftrightarrow P'$.
The last one follows {}from the third one 
by $\,\tilde{\;}\,$ and $P \leftrightarrow P', Q \leftrightarrow R$.
\end{proof}
\begin{lemma}\label{lemd:Kpro}
  \begin{align}
\label{DnAn-L1}
  &K_1 \cdots K_{n-1} 
    = \rho
      \begin{pmatrix}
L_n\Bigl[\begin{matrix} P^\prime & R \\ Q & P
                  \end{matrix}\Bigr] & \\
  & \Tilde{L}_n\Bigl[\begin{matrix} P & Q \\ R & P^\prime
                          \end{matrix}\Bigr] 
\end{pmatrix} 
      \rho,
  \\
  \label{DnAn-L2}
  &K_{-n+1} \cdots K_{-1} 
    = \begin{pmatrix}
 {}^t\bar{L}_n \Bigl[\begin{matrix} P & R \\ Q & P^\prime
                          \end{matrix}\Bigr] & \\
  & {}^t\Tilde{\bar{L}}_n \Bigl[\begin{matrix} P^\prime & Q \\ R & P
                                  \end{matrix}\Bigr]
\end{pmatrix},
\end{align}
where $\rho \in {\rm End}(V)$ denotes the  
interchange $v_n \leftrightarrow v_{-n}$.
\end{lemma}
\begin{proof}
We use induction on $n$.
The $n=3$ case is checked by a direct calculation.
Assume \eqref{DnAn-L1} and \eqref{DnAn-L2} are fulfilled 
up to $n$.
Then the left hand side of \eqref{DnAn-L1} for $n+1$ is 
\begin{align*}
  &K_1 K_2 \cdots K_{n} 
  \nonumber \\ 
  &~~ =
  \begin{pmatrix} 
  P_1^\prime & & & R_1 & & \\
  & \openone_{n-1}\\
  & & P_1^\prime & & & R_1 \\
  Q_1 & & & P_1 & & \\
  & & & & \openone_{n-1}\\
  & & Q_1 & & & P_1 
  \end{pmatrix} 
  \rho
  \begin{pmatrix}
  1 \\
  & L_n^+ \Bigl[\begin{matrix} P^\prime & R \\ Q & P
                  \end{matrix}\Bigr]\\ 
  & & \Tilde{L}_n^+ \Bigl[\begin{matrix} P & Q \\ R & P^\prime
                  \end{matrix}\Bigr]\\   
  & & & 1
  \end{pmatrix}
  \rho
  \nonumber \\ 
  & ~~ =  \rho
  \begin{pmatrix} 
  P_1^\prime & & R_1 & & & \\
  & \openone_{n-1}\\
  Q_1 & & P_1 & & & \\
  & & & P_1^\prime & & R_1 \\
  & & & & \openone_{n-1}\\
  & & & Q_1 & & P_1 
  \end{pmatrix} 
  \begin{pmatrix}
  1 \\
  & L_n^+ \Bigl[\begin{matrix} P^\prime & R \\ Q & P
                  \end{matrix}\Bigr]\\ 
  & & \Tilde{L}_n^+\Bigl[\begin{matrix} P & Q \\ R & P^\prime
                  \end{matrix}\Bigr]\\   
  & & & 1
  \end{pmatrix}
  \rho. 
\end{align*}  
Owing to the first two relations in Lemma \ref{lemd:ind},
this coincides with the right hand side of 
(\ref{DnAn-L1}) for $n+1$.
Similarly the induction assumption 
leads to the following expression for the 
left hand side of \eqref{DnAn-L2} 
for $n+1$:
\begin{align*}
  &K_{-n} K_{-n+1} \cdots K_{-1} 
  \nonumber \\ 
  &~~ =
  \begin{pmatrix}
  1 \\
  & ^t\bar{L}_n^+ \Bigl[\begin{matrix} P & R \\ Q & P^\prime
                  \end{matrix}\Bigr]\\ 
  & & ^t\Tilde{\bar{L}}_n^+ \Bigl[\begin{matrix} P^\prime & Q \\ R & P
                  \end{matrix}\Bigr]\\   
  & & & 1
  \end{pmatrix}
  \begin{pmatrix} 
  P_{-1}& & Q_{-1}\\
  & \openone_{n-1}  \\
  R_{-1} & & P_{-1}^\prime\\
  & & & P_{-1} & & Q_{-1}\\
  & & & & \openone_{n-1}\\
  & & & R_{-1} & &  P_{-1}^\prime 
  \end{pmatrix}.
\end{align*}  
Again the product can be computed by using the latter two relations 
in Lemma \ref{lemd:ind}, yielding the right hand side of 
(\ref{DnAn-L2}) for $n+1$. This completes the induction.
\end{proof}
\noindent
{\em Proof of Proposition \ref{prd:facL}}
The product $K_{-n+1} \cdots K_{-1} D(z) K_{1} \cdots K_{n-1}$ 
can be calculated by using 
Lemma \ref{lemd:Kpro}, (\ref{eqd:dd}) and (\ref{eqa:L-elements}).
The result agrees with the $L(z)$ defined in section \ref{subsec:Ld}.
\vspace{-0.57cm}\begin{flushright}$\square$\end{flushright}

\section*{Acknowledgements}
The authors thank Taichiro Takagi and Yasuhiko Yamada for discussion.
A.K. thanks Murray Batchelor, Vladimir Bazhanov, Vladimir Mangazeev and 
Sergey Sergeev for a warm hospitality at the Australian National University 
during his stay in March 2004.
A.K. and M.O. are partially supported by Grand-in-Aid for Scientific 
Research JSPS No.15540363 and No.14540026, respectively 
from Ministry of Education, Culture, 
Sports, Science and Technology of Japan.

\end{document}